\documentclass[12pt,letterpaper]{article}

\usepackage{geometry}
\pdfoutput=1 

\usepackage[utf8]{inputenc}
\usepackage{cite}
\usepackage{nameref,hyperref}
\usepackage[right]{lineno}
\usepackage{microtype}
\DisableLigatures[f]{encoding = *, family = * }

\usepackage{changepage}
\usepackage[aboveskip=1pt,labelfont=bf,labelsep=period,singlelinecheck=off]{caption}
\makeatletter
\renewcommand{\@biblabel}[1]{\quad#1.}
\makeatother

\usepackage{lastpage,fancyhdr,graphicx}
\usepackage{epstopdf}
\usepackage{natbib}
\usepackage{amsmath}
\fancyheadoffset[L]{2.25in}
\fancyfootoffset[L]{2.25in}

\usepackage{color}

\definecolor{Gray}{gray}{.25}

\usepackage{graphicx}

\usepackage{wrapfig}
\usepackage[pscoord]{eso-pic}
\usepackage[fulladjust]{marginnote}
\reversemarginpar

\usepackage{algorithm} 
\usepackage[noend]{algpseudocode}
\usepackage{enumitem}
\makeatletter
\def\BState{\State\hskip-\ALG@thistlm}
\makeatother

\begin{document}
\vspace*{0.35in}
\begin{flushleft}
{\Large
\textbf\newline{Analysis of association football playing styles: an innovative method to cluster networks}
}
\newline
\\
Jacopo Diquigiovanni\textsuperscript{1,*},
 Bruno Scarpa\textsuperscript{2}
\\
\bigskip
\bf{1} Department of Statistical Sciences, University of Padova, Italy
\\
\bf{2} Department of Statistical Sciences, University of Padova, Italy
\\
\bigskip
* jacopodiquigiovanni93@gmail.com

\end{flushleft}

\section*{Abstract}
In this work we develop an innovative hierarchical clustering method to divide a sample of undirected weighted networks into groups. The methodology consists of two phases: the first phase is aimed at putting the single networks in a broader framework by including the characteristics of the population in the data, while the second phase creates a subdivision of the sample on the basis of the similarity between the community structures of the processed networks. Starting from the representation of the team's playing style as a network, we apply the method to group the Italian Serie A teams' performances and consequently detect the main 15 tactics shown during the 2015-2016 season. The information obtained is used to verify the effect of the styles of play on the number of goals scored, and we prove the key role of one of them by implementing an extension of the Dixon and Coles model \citep{dixon1997modelling}.

\section{Introduction}
In recent years the exponential growth of available data and the progressive opening up to 'statistical culture' by football operators have led to the 
development of several studies in this field \citep{anderson2013numbers}. The unique nature of association football means that simple statistics do not allow a detailed measurement of team performances: in fact, the signal contained in the data is hardly extractable since many accidental events may increase the noise, especially if data consist only of low counts such as the number of goals scored during the match \citep{pena2012network}.

As a consequence, the \emph{playing style} (or \emph{tactic}) is certainly an element of primary importance in this sport \citep{pena2015can}. Generally speaking, it is possible to define it as a combination of two different aspects: the \textit{offensive} playing style, defined as the way a team moves the ball on the pitch, and the \textit{defensive} playing style, defined as the way a team defends against the way the opposing team moves the ball on the pitch. In this work we focus on the former, and so  
later in the discussion the term 'playing style' will be used only to refer to the offensive playing style. Over time, some studies have addressed this issue: for example, \citet{hughes2005analysis} compare the length of passing sequences, 
\citet{gyarmati2014searching} analyse the passing structures to find similarities and differences between teams, \citet{cintia2015harsh} and \citet{cintia2015network}  
 propose a pass-based performance indicator to predict the outcome of the matches and \citet{pena2014markovian} studies the possession by using a 
Markov process.





In statistical terms, the playing style could be adequately represented by a weighted network \citep{brandt2015graph, clemente2015social}. Most of the papers in this field consider a team's players as 
nodes and the passes between them as edges to extract information on the tactics \citep{pena2012network, clemente2015using,
 pina2017network}. Unfortunately, this formulation makes comparisons between different playing styles not immediate as the players vary from squad to squad. 
In view of this, this paper considers networks whose nodes are different areas of the pitch 
and whose edges describe the movements of the ball between these areas; in so doing the playing styles are properly summarised and it is also possible to make comparisons because the pitch is obviously shared between the teams. 

The aim of the paper is to group the networks in order to compare the different styles played by football teams during the season. In order to do this, we
propose a novel clustering method for network data, the aim of which is to divide the $n$ weighted networks into clusters according to a specific criterion. The authors are not aware of specific methods to detect clusters of networks, although some models \citep[e.g., the bayesian model proposed by][for binary networks]{durante2016nonparametric} can be applied for this purpose, so the present paper suggests an initial proposal  in this field of research. 
In this scenario, a couple of aspects should be considered: the network is, in fact, a set of connected nodes, but also a 
unit drawn from a more general population of networks. In view of the above, after the preprocessing phase that puts the networks in a sample framework, the 
method defines the similarity between networks on the basis of the similarity between community structures detected by the Louvain method \citep{blondel2008fast}. The 
methodology has a hierarchical agglomerative structure: from the initial situation in which each unit is assigned to a separate group, the process creates a sequence 
of nested subdivisions of data.

The method, by clustering the performances of the teams as networks, identifies the main styles of play, and this information is used to verify whether and how a 
team's playing style affects the number of goals scored by implementing an extension of the Dixon and Coles model \citep{dixon1997modelling}.

The paper is organised as follows: in Section 2 we create the networks from the available data; 
in Section 3 the clustering
method is developed; in Section 4 the method is applied to the analysis of tactics and in Section 5 an overview of the main results is provided.

\section{Data}

\label{sectiondata}
Different approaches are available to analyze association football tactics \citep{anderson2013numbers}. Among them, the representation of the playing style as a network is particularly useful  \citep{pena2012network}. In this work, for each combination of match and team (i.e., for each of the two teams playing a specific match) the playing style is represented by a network  whose nodes are the different areas into which the pitch is divided  and whose edges describe the movements of the ball between these areas: this specification is preferable to the one that considers the players as nodes and the passes between them as edges \citep{clemente2015using} because it allows immediate comparison between different tactics since the nodes (i.e. the areas of the pitch) are shared between the teams.
Specifically, we consider weighted undirected networks \citep{newman2010networks}. 

The first factor to consider is the shape and the number of areas $K$ identifying the different nodes. As regards the first aspect,  for the sake of simplicity the areas  are obtained by dividing both the length and the width of the pitch into equal segments \citep{borrie2002temporal, narizuka2014statistical}. 
As regards the second aspect, in a preliminary study we have compared the partitions obtained by the clustering method presented in Section \ref{sectionproposedmethodology} with various values of $K$ ($K=9$, $K=12$, $K=18$), finding deep differences between the three scenarios. This evidence represents a critical issue since we are not aware of specific information to set $K$: consequently, the choice of $K$ is based purely on subjective considerations, and we decide to choose $K=9$ because it is associated with the best known subdivision of the  length of the pitch (defense zone - midfield zone - attack zone) and with the best known suddivision of the width of the pitch (right zone - central zone - left zone).

The available data - provided by InStat (\url{http://instatfootball.com/}) - refer to the 380 matches of the Italian Serie A TIM 2015-2016 season; since each match is made up of two distinct networks - one for the home team and one for the away team -, the sample size is $n=760$.

For each combination of match and team, the available data detect spatial coordinates $(x,y)$ of specific plays made by the team's players during the match, with the $x$ position along the horizontal axis - i.e., the length of the pitch - and the $y$ position along the vertical axis - i.e., the width of the pitch -. These plays are divided into 12 categories, which can be summarised in four macro categories:

\begin{itemize}
\item Passes: the starting position of a pass. It includes: accurate passes, inaccurate passes, assists, accurate crosses and inaccurate crosses.
\item Dribbling: the starting position of a dribble. It includes: successful dribbles and unsuccessful dribbles.
\item Tackles: the position of a tackle. It includes: successful tackles, unsuccessful tackles, challenges won and challenges lost.
\item Shots: the position of the shot.
\end{itemize}
Starting from the spatial coordinates, each play is assigned to one of the nine areas, labeled by the numbers from 1 to 9 (see Figure 1).
Hence the weight of the edge linking node $i$ and node $j$ is represented by the number of pairs of consecutive plays made by the players of the team without
interruption from the players of the opposing team and which take place in areas $i$ and $j$ respectively. If an event is not preceded and followed by a teammate's play, 
this is ignored as an isolated event; moreover, the proposed structure allows the creation of \emph{self-loops} when $i=j$. The main characteristics of the networks are summarised in Appendix A. 


Naturally, the networks are only an approximation of the playing styles: the data do not consider the movements of the players when in possession
of the ball. As a consequence, the ball is considered to have passed only through the $i$ and $j$ areas in the time between two consecutive events which have occurred in 
those areas. Based purely on football considerations, one is justified in expecting the teams whose playing style is strongly characterised by the movements with the
ball by the players to be different from the teams where this is more limited; the level of accuracy affecting the approximation therefore changes according to the  unit considered. Nevertheless, the study of the number of plays recorded by our data during the season is informative with regard to this loss of information: the average number of events during the
380 matches, taking into account the effective playing time of an Italian football match \citep{Santinigoal}, is 23.96 per minute ($sd=1.62$), an amount that the authors consider satisfactory.

\begin{figure}
\begin{center}
\includegraphics[width=1\textwidth,height=6.8cm]{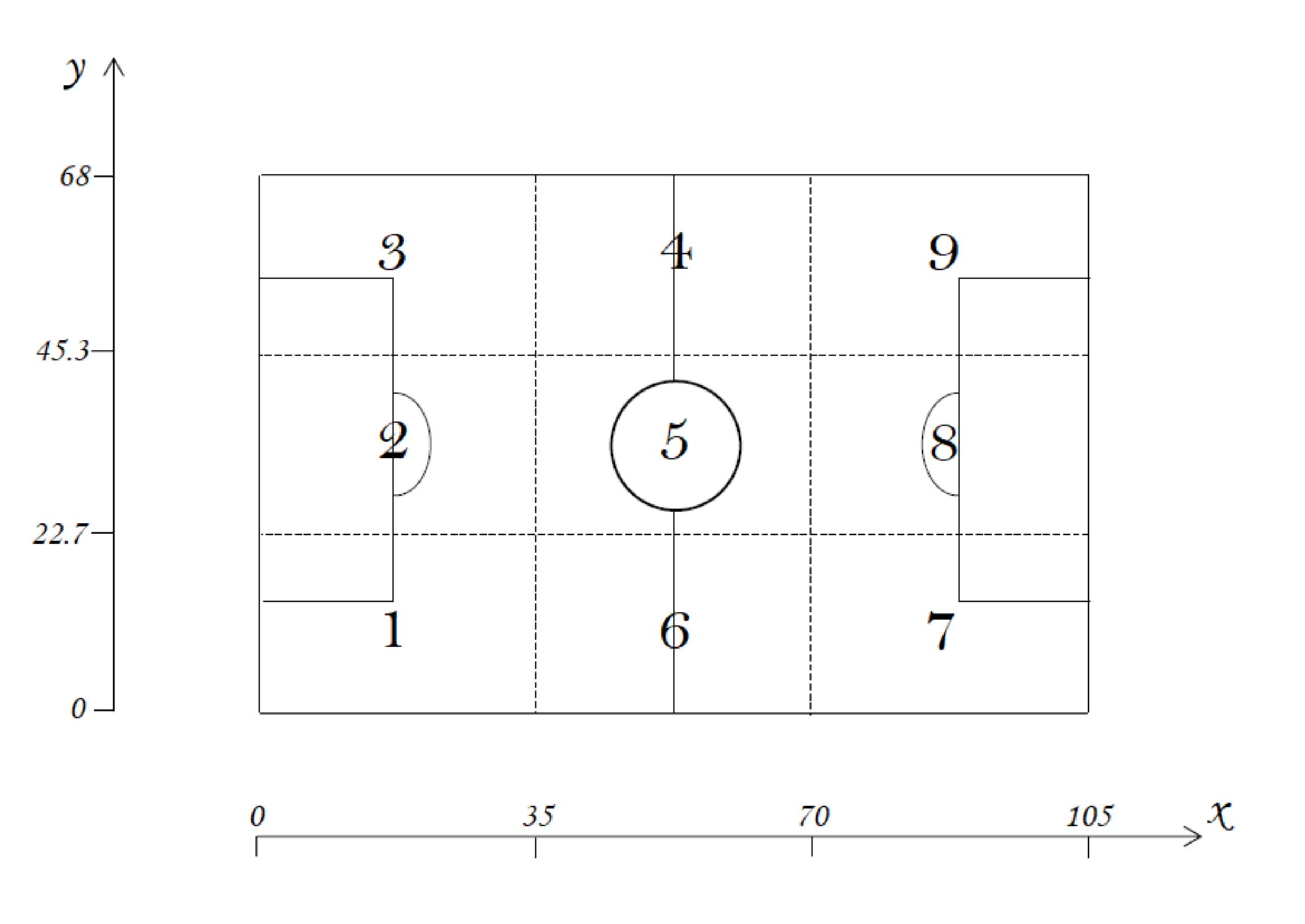}
\caption{Division of the pitch into nine areas. The pitch measurements are 105 $\times$ 68 m: the areas are obtained by dividing both the length and the width of
the pitch into three equal segments. The teams attack from left to right. 
 }
\label{IMMsuddivisionecampodacalcio}
\end{center}
\end{figure}

\section{Proposed methodology}
\label{sectionproposedmethodology}
In this section we propose a clustering method for network data, the aim of which is to divide the $n$ undirected weighted networks into groups according to a specific criterion: the only constraint required is that the statistical units share the set of nodes $\mathcal{K}=\{k_1, k_2,\dots,k_K\}$.

The procedure 
has a hierarchical structure: in contrast to the partitioning approach, in which at each step the statistical units are reallocated in a number of clusters fixed a priori, the hierarchical approach creates nested subdivisions of data \citep[for this and further differences see][]{kaufman1990finding}. The method is agglomerative: starting from an initial state in which the number of groups is equal to the number of networks, the algorithm continues until all the units belong to the same group. As a consequence, the partition at step $h$ can be obtained from the previous one by merging two clusters. We use an \emph{ad hoc} measure of similarity to compare the groups, and the number of clusters - if it is not known a priori - is defined by the analysis of the
corresponding dendrogram. In addition, the methodology consists of two phases: the preprocessing phase, that requires the specification of a threshold, and the merging phase. A simulation study is provided in Appendix B.



\subsection{Similarity between networks}
\label{sectionsimilaritybetweennetworks}
Focusing  on a single network, the core of the cluster analysis is represented by a set of procedures known as \emph{analysis of community
structure}; it is essential in many applications to detect communities of nodes with frequent connections within communities and sparse connections between 
communities in order to find latent relationships in the data \citep{newman2010networkscap11}. In this context, the Louvain method \citep{blondel2008fast} is certainly a widely used approach, and it represents the starting point of this paper. 

Focusing now on a sample of $n$ networks with $K$ shared nodes, a \emph{naive} clustering method consists of detecting the $n$ community structures and merging, step by step, the networks whose partitions are most similar. The limit of this approach is clear, since every unit is taken individually regardless of its close relationship with the rest of the data. In this scenario, a couple of aspects should be taken into account: the network is, in fact, 
considered on the one hand as a set of connected nodes, and on the other as a unit drawn from a more general population of networks.  Regarding the latter, it is essential to stress that:
\begin{itemize}
\item The average number of connections can vary between the networks, with some networks sparsely connected and others densely connected.
\item The weight of the edges can vary within the networks. 
\item The weight of a specific edge can vary between the networks. As a consequence, the community structures can differ considerably. 

\end{itemize}
In view of the above, it is first of all necessary to define the concept of \emph{similarity} between networks. The method considers the community structures detected by the Louvain method after including the population's characteristics in the units: in other words, the similarity between networks $R_{t}, R_{s}$ is determined according to the similarity of the community structures of the networks $R^{'}_{t}, R^{'}_{s}$,  with the network $R^{'}$ representing appropriate processing of starting network $R$.

\subsection{First phase: preprocessing of the networks}
\label{firstphase}
The procedure to obtain $R^{'}$ is described below. Considering the sequence of weights linked to the edge connecting nodes $i$ and $j$ in $n$ networks  $w_{ij}=(w_{ij}^1,\dots,w_{ij}^n)$, we define:
\begin{equation}
b_{ij}=\max(w_{ij}^1,\dots,w_{ij}^n)  \quad \quad a_{ij}=\min(w_{ij}^1,\dots,w_{ij}^n)  
\label{formulamassimominimo}
\end{equation}
with $i,j=1,\dots,K$ and then \emph{normalize} vector $w_{ij}$:
\begin{equation}
u^{t}_{ij}=\begin{cases}
\frac{w_{ij}^t-a_{ij}}{b_{ij}-a_{ij}} \in [0,1] & \text{if $a_{ij} \neq b_{ij}, \quad t=1,\dots,n$}\\ 
0.5 & \text{if $a_{ij} = b_{ij}, \quad t=1,\dots,n$} 
\end{cases}
\label{normalizzazione}
\end{equation}
For the sake of simplicity, if $a_{ij}=b_{ij}$ the value $u^t_{ij}$ is set equal to the expected value of a random variable with continuous uniform distribution $U(0,1)$. Therefore a high value of the normalized quantity means the relationship between the pair of nodes is numerically significant compared to that observed in the other units, while a low value means the relationship is numerically inferior to the others. The decision to change the values of $a$ and $b$ according to the pair of nodes considered allows the distributive differences between the sequences of weights $w_{ij}$ to be included in the analysis, an essential topic which has already been discussed. 

Unfortunately, the normalization is functional but not comprehensive because another problem commonly arises in this context. Let us consider the following example: \begin{figure}
\begin{center}
\includegraphics[width=0.75\textwidth,height=6cm]{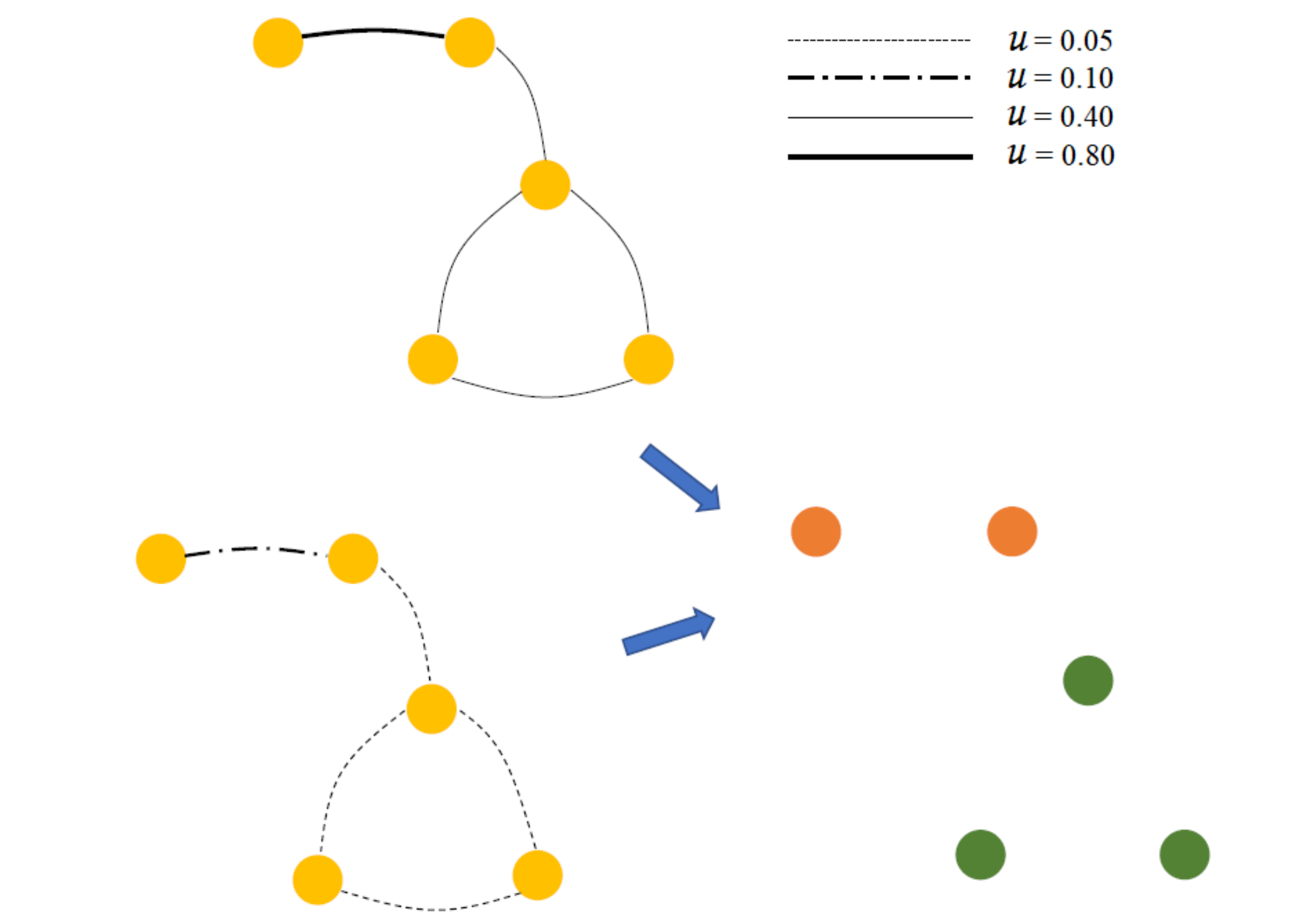}
\caption{Example of problem in the application of Louvain method.}
\label{IMMproblematicalouvain}
\end{center}
\end{figure}let $K=5$ and $R_1$ be a network in which all the normalized weights are extremely low $u_{1,l}=0.05$ except for the edge between node 1 and node 2 which weighs twice as much as the previous ones $u_{1,h}=0.10$ (see the bottom left of Figure \ref{IMMproblematicalouvain}). Let $R_2$ also be a network in which all the normalized weights are definitely higher $u_{2,l}=0.40$ and, as in $R_1$, the edge between the first two nodes weighs twice as much ($u_{2,h}=0.8$, see the up of Figure \ref{IMMproblematicalouvain}). The community structures of the two networks are identical (see the bottom right of Figure \ref{IMMproblematicalouvain}): in each case the first two nodes belong to a cluster distinct from the one containing the remaining nodes and consequently the two units have maximum similarity, assuming of course the use of a measure of similarity that assigns maximum value if the partitions are identical. However in the first case the group including nodes 1 and 2, considering the particularly low normalized quantity $u_{1,h}$, is found only because the other 
weights of the network are even smaller; evidently in this situation detecting this cluster is counterproductive as $u_{1,h}$ is much lower than the other connections observed in the sample between the first pair of nodes. The reason for this critical issue lies in the use of the Louvain method which, by construction, compares the edges within the network regardless of the broader framework of this paper.

For this reason, a possible approach to limit the problem consists of introducing a threshold $s \in [0,1]$. Starting from the normalized weight $u_{ij}^t$ (\ref{normalizzazione}) we obtain the value $w^{ ' t}_{ij}$ defined as follows:
\begin{equation}
w^{ ' t}_{ij}=\begin{cases}
u_{ij}^t & \text{if $u_{ij}^t > s$} \qquad t=1,\dots,n\\ 
0 & \text{if $u_{ij}^t \leq s$} \qquad t=1,\dots,n
\label{pesinormalizzatiSunico}
\end{cases} 
\end{equation}
Hence the processed network $R'$ has the same nodes as starting network $R$, whereas the weights are $w^{'}_{ij}$ instead of $w_{ij}$. The choice of the threshold plays a central role: indeed numerically formula (\ref{pesinormalizzatiSunico}) does not alter the weights greater than $s$, while it sets all the other ones to zero. In view of the above, on the one hand a high threshold value ensures that only the communities characterised by remarkable connections for the population are detected; on the other hand it does not discriminate between weights less than or equal to $s$. Conversely, a near-zero threshold value involves the problem of the example provided but does not involve loss of information.

At the interpretative level, the two proposed adjustments - normalization first, followed by introduction of threshold - relate to two complementary and jointly necessary aspects: the aim of the former is to unify the range of observable values so as not to penalise the connections between pairs of nodes that are sparser than others due to the population's characteristics; conversely, the aim of the latter is to curb the negative implications of the Louvain method due to its application \emph{network by network}.

The choice of $s$ represents the last fundamental step in the preprocessing phase. Considering the possible distributive differences between the sequences of weights $w_{ij}$, 
it is reasonable to vary $s$ according to the value of the couple $(i,j)$. We therefore define: 
\begin{equation}
s_{ij}=q ( u_{ij},\alpha) \qquad \qquad i,j=1,\dots,K
\label{sogliavariabile}
\end{equation}
with  $u_{ij}=(u_{ij}^1,\dots,u_{ij}^n)$ and $q(x,\alpha)$  quantile of order $\alpha$ for $x$. Formula (\ref{sogliavariabile}) appears to be a reasonable compromise: the threshold could vary depending on the distributive characteristics of the vector
 $u_{ij}$ despite the specification of a single scalar $\alpha$. Formally, formula (\ref{pesinormalizzatiSunico}) is updated as follows:
\begin{equation}
w^{ ' t}_{ij}=\begin{cases}
u_{ij}^t & \text{if $u_{ij}^t > s_{ij}$} \qquad t=1,\dots,n\\ 
0 & \text{if $u_{ij}^t \leq s_{ij}$} \qquad t=1,\dots,n
\end{cases} 
\label{pesinormalizzatiSvariabile}
\end{equation}
Since the cluster analysis is typically an unsupervised learning technique, frequently the choice of $\alpha$ can not be achieved by using a training set \citep[see, e.g., ][]{hastie2009elements}: as a consequence,  although a simulation study (see Appendix B) may provide some useful information, the value of the
threshold shall be selected on the basis of the analyst choice to place more emphasis on one of the two abovementioned aspects.


\subsection{Second phase: merging phase}
\label{paragrafofase2}
Starting from the new networks, it is possible to detect the $n$ community structures present in the sample. As the methodology would like to create
groups composed of networks with similar connections between nodes, before the application of the Louvain method the weight of every self-loop is set to zero. Indeed, two networks whose adjacency matrices share the entries outside the main diagonal could present deeply different community structures if the diagonal elements
are considered, an aspect that inevitably causes undesirable effects in the reliability of the clustering method.
Clearly, the interpretation of the community structures changes radically when considering the processed networks rather than the starting ones: any community denotes a relationship between nodes which is particularly surprising for the population and not necessarily identifiable by the analysis of the original units.

Given the nature of the methodology, it is now possible to justify the use of undirected networks since the utilisation of directed ones does not induce
improvements. Theoretically, the inclusion of the edges' direction  adds useful information because the role of connection $i \rightarrow j$ is different from that of connection $j \rightarrow i$.
However, the proposed method detects only the community structures that, by definition, do not have direction and therefore it cannot, by construction, capture this
aspect. As a consequence, for the sake of simplicity we have considered undirected networks.

The tool used to define the similarity between a pair of partitions is the Adjusted Rand Index \citep[or ARI, ][]{hubert1985comparing}, the most common corrected-for-chance
version of the Rand Index \citep{rand1971objective}. By calculating this index for each pair of units it is possible to create a \emph{similarity matrix} and so to
merge the groups using the UPGMA method \citep{sokal1958statistical}: the procedure ends when all the networks belong to the same cluster.

The choice of the number of groups $N_G$ represents the last fundamental step in the merging phase. In this context, the evaluation of the difference between
the maximum values of similarity (i.e., the maximum values of ARI) in two successive steps of the method is common practice for the hierarchical procedures \citep[see, e.g.,][]{azzalini2012data}: for example, a 
significant drop in similarity provides guidance that the clusters are too different to proceed with subsequent unions. Obviously, as often happens in this field, the 
 criterion cited can be accompanied by qualititative evaluations by the analyst.

\subsection{The algorithm}
A brief overview of the procedure is provided in Algorithm \ref{euclid}. Let $R=\{ R_1,\dots,R_n \}$  be a sample of $n$ undirected weighted networks, with $\mathcal{K}=\{k_1, k_2,\dots,k_K\}$ the  set of nodes shared between the networks. The value $w_{ij}^t$ is the weight of the edge connecting node $i$ and node $j$ in network $t$, with $t=1,\dots,n$ e $i,j=1,\dots,K$. The possible absence of connection between a couple of nodes involves a zero weight linked to that edge.
Let  $\alpha_0$ also be the selected value for $\alpha$. The algorithm is structured as follows: in the first phase, calculate quantity $w_{ij}^{'t}$ by the normalization of data and the introduction of the threshold (line \ref{riga1}), therefore obtain the new sample $R^{'}=\{R_1^{'},\dots,R_n^{'}\}$ (line \ref{riga2}). In the second phase, set the self-loops of the networks to zero (line \ref{riga3}), detect the community structure for every network using the Louvain method (line \ref{riga4}) and create the similarity matrix (line \ref{riga5}). Finally,
two steps proceed iteratively: obtain the pair of groups with maximum similarity using the UPGMA method (if it is not unique, the pair is randomly chosen from those that maximize the ARI) and merge the pair of groups selected at the previous step (lines \ref{riga9}-\ref{riga14}).
The procedure ends when all the networks belong to the same group.


\begin{algorithm}
\caption{Clustering method for sample of networks}\label{euclid}
\begin{algorithmic}[1]
\State \textbf{Start:}  each network is assigned to a separate group
\Procedure{First phase}{}
\For { \textit{t} \text{in $1:n$}, \textit{i} \text{in $1:K$}, \textit{j} \text{in $1:K$} }  $ w^{ ' t}_{ij} \gets$ Formulas (\ref{formulamassimominimo}),(\ref{normalizzazione}),(\ref{sogliavariabile}),(\ref{pesinormalizzatiSvariabile})              \label{riga1}                       

\EndFor
\State \textbf{for} \textit{t} \text{in $1:n$} \textbf{obtain} $R^{'}_{t}$ new network \label{riga2}, with $\mathcal{K}$ as set of nodes and weights $w_{ij}^{'t}$  
\EndProcedure
\Procedure{Second phase}{}
\For { \textit{i} \text{in $1:K$} }                        $ w^{ ' }_{ii} \gets \textbf{0}$ \label{riga3}
\EndFor
\For { \textit{t} \text{in $1:n$} }                        $ c^{ t} \gets$ \text{community structure of $R_t^{'}$ by Louvain Method} \label{riga4}
\EndFor
\For { \text{$t_1$ in $1:(n-1)$}, \text{$t_2$ in $(t_1+1):n$}  } $s(R^{'}_{t_1},R^{'}_{t_2}) \gets ARI(c^{t_1},c^{t_2})$ \label{riga5}
\EndFor

\While{number of groups $\neq 1$} \label{riga9}
\If{pairs of groups with max similarity using the UPGMA method $>1$}
\State select a pair of groups randomly from those that maximize the ARI
\Else \{the pair of groups with max similarity is unique\}
\State select it \textbf{end if}

\EndIf

\State Merge the pair of groups selected \textbf{end while} \label{riga14}
\EndWhile


\EndProcedure

\end{algorithmic}
\end{algorithm}

\section{Analysis of playing styles}

\subsection{Application of the method}
The first aspect to consider in the application of the clustering method 
is the choice of the threshold $s$: since we do not have either a training set or external information for set $\alpha$, it is based on the balance between the two contrasting features presented in Section \ref{firstphase}. The selected value for $\alpha$ is 0.95.
Such a high threshold is aimed at sacrificing strong relationships detected in the data 
in order to obtain community structures characterised by incredibly deep connections between nodes: 
in so doing, the problem related to the  \emph{network by network} application of the Louvain method is considerably reduced. On the other hand, the methodology does not consider the obvious heterogeneity in the large set of values which are less than or equal to the threshold.

Once the method is applied, the analysis of the dendrogram provides useful indications about the procedure: in the first phase the examination of the number of mergers with maximum similarity,
i.e. the number of steps whose maximum ARI is equal to one, is informative. The value found is 519, and the corresponding partition consists of 59 clusters with more than
one unit and 182 singletons. This subdivision shows all the 241 different playing styles of the Italian Serie A TIM 2015-2016 season: the
following mergers group these tactics proposing clusters characterised by increasing heterogeneity.

The final selection of the number of groups is based on the balance of statistical criteria and practical necessities. As explained in Section \ref{paragrafofase2}, the evaluation of the difference between 
the maximum values of similarity in two successive steps of the method is common practice when choosing $N_G$ which, on the other hand, should not be too high to allow the analysis of the partition obtained. 
In the case of 26 groups, the maximum ARI is less than 0.10 (ARI$=0.092$), a decidedly low value: as
a consequence, the selected value is $N_G=15$ in order, at least, to obtain a manageable number of clusters. The detailed description of the groups is provided in Appendix C.

\subsection{Modelling final scores}
\label{modellingfinalscores}
Modelling the final result of a football match represents a field of deep interest in sports analysis. Moving from the paper written by \citet{maher1982modelling},  several 
authors have proposed increasingly complex models: \citet{dixon1997modelling}, \citet{baio2010bayesian} and \citet{koopman2015dynamic} are only a few examples. 
The aim of this section is to verify whether and how a team's playing style affects the number of goals scored. In a preliminary study we have considered a model which is able to bring together the main features of Maher's approach and
the information about playing styles, i.e. the Poisson regression with canonical link function. 
This simple model seems to suggest that the on-the-wings playing style (see Appendix C)  affects the number of goals scored (for all the details, contact J.Diquigiovanni). This explorative model undoubtedly represents an oversimplification of 
the phenomenon, and it is unnecessary to focus on it: so as to verify the role of the group 15, an extension of the Dixon and Coles model is implemented as follows.

Let $X_k \sim Poisson(\lambda_k)$ be the number of goals scored by the home team in match $k$ with $k=1,\dots,380$  and let $Y_k\sim Poisson(\mu_k)$ be the number of goals scored by the away 
team on the same occasion;  therefore a home team $i$ and an away team $j$ are implicitly assigned to every match $k$ with $1 \leq i \neq j \leq 20$.
For convenience, the matches are considered chronologically and the season is divided into a series of half-weekly time points.
\citet{dixon1997modelling} propose the use of a `pseudolikelihood'  $L_t(\alpha_i,\beta_i,\rho,\gamma; i=1,\dots,20)$ for each time point $t$, with:
\begin{equation}
\log(\lambda_k)=\gamma + \alpha_{i(k)} + \beta_{j(k)} \quad \quad \log(\mu_k)=\alpha_{j(k)} + \beta_{i(k)} 
\label{ciao1ciao2}
\end{equation}
and with $\gamma$ parameter which allows for the home effect, $\alpha_{i(k)}$,$\beta_{i(k)}$ parameters which measure respectively the attack and defence rates of the teams, $i(k)$, $j(k)$ indices which denote
the home and away teams. 


So as to verify the effect of the 
on-the-wings playing style on the number of goals scored, the suggested adjustment is conceptually simple. Starting from equations (\ref{ciao1ciao2}) , we
obtain:
\begin{equation*}
\log(\lambda_k)=\gamma + \alpha_{i(k)} + \beta_{j(k)} + \delta c_{i(k)} \quad \quad \log(\mu_k)=\alpha_{j(k)} + \beta_{i(k)} + \delta c_{j(k)}
\label{lambdamumodificato}
\end{equation*}

with  $c_{i(k)}=1$  if the playing style of team $i$ during match $k$ is assigned to group 15, 0 otherwise; $\delta$ effect of this tactic on the number of goals scored. In addition, for convenience, the constraint $-\sum_{i=1}^{19} \alpha_i=\alpha_{20}$ is used instead of the one used by Dixon and Coles 
$\sum_{i=1}^{20} \alpha_i=20$.

In contrast to the original model, the time points are not homogeneously distributed over the season, but the specific day of the year on which the match $k$ takes place is considered. 
This adjustment is due to the fact that nowadays, unlike when Dixon and Coles carried out the study, 
the teams play almost every day of the week and so a more precise subdivision of the season is recommended. 

Since the quantity of interest is $\delta$, the inference focuses only on this parameter. Starting from the composite profile likelihood \citep[see, e.g.,][]{libropseudoveros} defined as:
\begin{equation*}
L_t^P(\delta)=L_t(\delta, \hat{\alpha}_{i_{\delta}},\hat{\beta}_{i_{\delta}},\hat{\rho}_{\delta},\hat{\gamma}_{\delta})=L_t(\delta,\hat{\theta}_{\delta})
\end{equation*}
with $\theta=(\alpha_i,\beta_i,\rho,\gamma)$ set of nuisance parameters, it is possible to construct confidence intervals on the basis of the composite likelihood ratio statistic \citep{varin2011overview}. Let us define $\psi=(\delta,\theta)$, $u_t(\psi)=\nabla_{\psi} \log L_t(\psi)=\nabla_{\psi} l_t(\psi)$, the sensitivity matrix at time $t$ as $H_t(\psi)=E_{\psi} \{- \nabla_{\psi} u_t(\psi)\}$, the variability matrix at time $t$ as $J_t(\psi)= var_{\psi} \{ u_t(\psi)\}$, the Godambe information matrix \citep{godambe1960optimum} at time $t$ as $G_t(\psi)= H_t(\psi) J_t(\psi)^{-1} H_t(\psi)$ and with  $H_t^{\delta \delta}$ ($J_t^{\delta \delta}, G_t^{\delta \delta}$ respectively) the inverse of $H_t(\psi)$ ($J_t(\psi), G_t(\psi)$ respectively) pertaining to $\delta$.
Therefore according to the asymptotic result proposed by \citet{satterthwaite1946approximate}, the confidence intervals are defined as follows:
\begin{equation*}
\biggl\{ \delta : \frac{2   \bigl( l_t^P (\hat{\delta}) - l_t^P (\delta) \bigl)}{ (H_t^{\delta \delta})^{-1} G_t^{\delta \delta}} < \chi^2_{1;{1-\alpha}} \biggr\}
\end{equation*}
Since there is only one parameter of interest, the solution provided corresponds to that established by \citet{geys1999pseudolikelihood}.
From a computational point of view, \citet{varin2011overview} propose the following quantities to estimate the sensitivity and variability matrix when the sample size is large:
\begin{equation*}
\hat{H}_t(\psi)= -\frac{1}{|A_t|} \sum_{k \in A_t} \nabla u^k (\hat{\psi}_{CL}) \quad \quad \hat{J}_t(\psi)= \frac{1}{|A_t|} \sum_{k \in A_t}  u^k (\hat{\psi}_{CL}) u^k (\hat{\psi}_{CL})^T  
\end{equation*}
with $A_t= \{ k: t_k < t \}$, $|A_t|$ cardinality of $A_t$, $u^k (\psi)$ score associated with log-likelihood term $l^k(\psi)$ such that $l_t(\psi)=\sum_{k \in A_t} l^k(\psi)$, $\hat{\psi}_{CL}$  maximum composite likelihood estimate.

As a consequence, in order to have a satisfactory sample size, the confidence intervals are computed from February 14th, 2016 ($t=61$), whereas the start of the season is used to optimize the choice of the parameter $\xi$ through the procedure described by Dixon and Coles; the selected value is 0.003 and this remains constant for the whole season. In addition, the last five matches of each team are not considered since the achieving of the team goal - whether this is to survive relegation or to win the league -  may involve a lack of effort affecting the performances.
As a result, the confidence intervals are computed to April 17th, 2016 ($t=84$) for an overall amount of 24 distinct matchdays: note that, by construction,  the progressive increase in the data considered causes a decrease in the width of the intervals, with obvious improvements in inference.

\begin{figure}
\begin{center}
\includegraphics[width=1\textwidth,height=8.5cm]{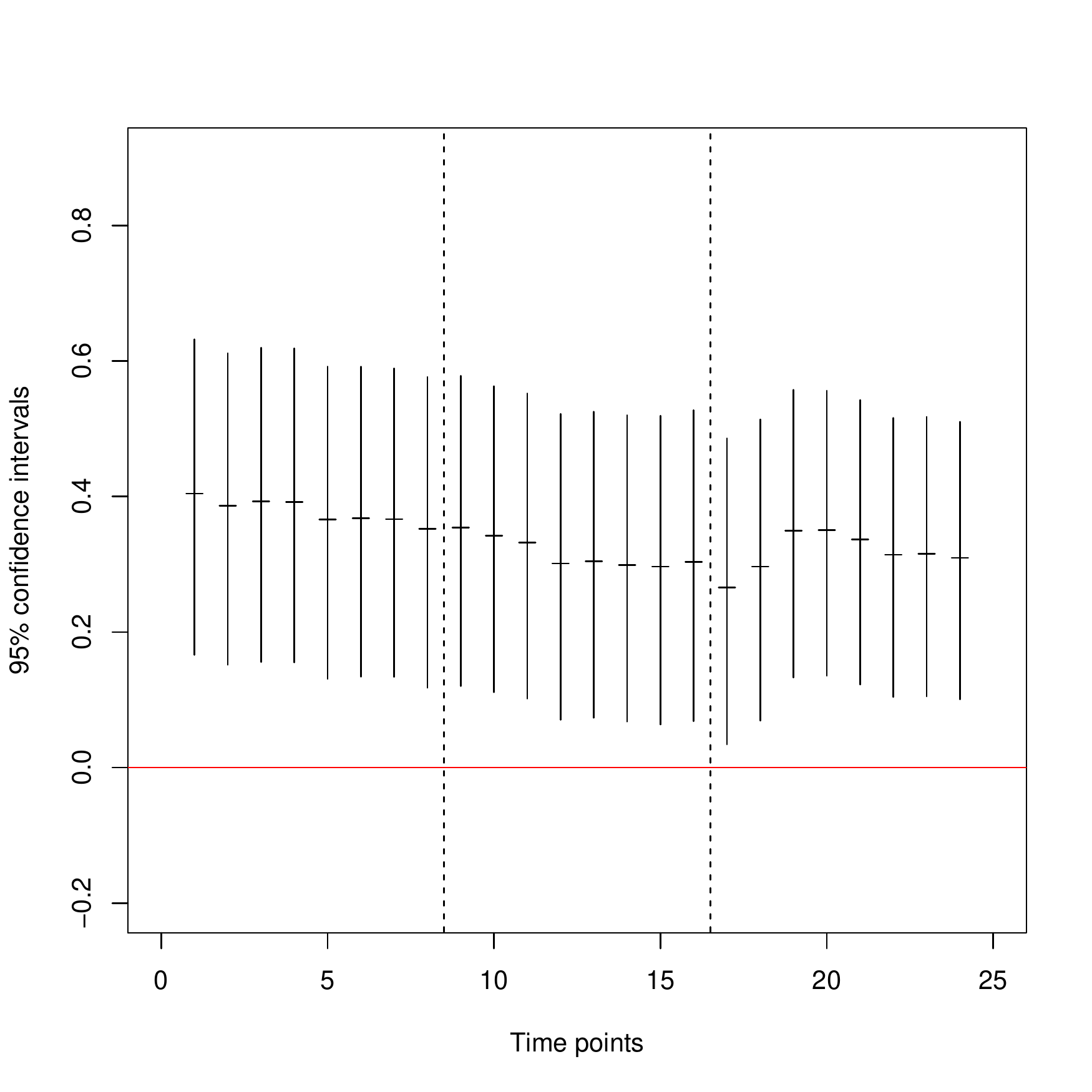} 
\caption{95\% confidence intervals and point estimates (horizontal dashes) for parameter $\delta$. The dashed lines divide the time points into months (February-March-April).}
\label{IMM5}
\end{center}
\end{figure}

Figure \ref{IMM5}
 shows the 95\% confidence intervals. The results should be approached with particular caution since the composite profile likelihoods are not independent: nevertheless, the evidence 
in the period considered involves fairly inferential conclusions. Indeed, all the intervals include only positive values: the on-the-wings playing style seems to affect the number of goals scored positively. While maintaining its typical oscillatory trend, the sequence of point estimates is a decreasing one: a possible explanation is that, after an initial adjustment period, the teams take technical-tactical precautions to limit this tactic, causing a decrease in its attacking efficiency.

\section{Discussion}
We have developed an innovative hierarchical clustering method to divide a sample of undirected weighted networks into clusters. The procedure represents a flexible
solution as the choice of the threshold $s$ allows the characteristics of the population to be managed on a case-by-case basis, and so a varied set of scenarios can
be properly addressed. On the other hand, the balance of the trade-off between the loss of information and the reliability of community structures is a potential limit in an
unsupervised context: as illustrated through simulations, a good choice of $s$ produces great results, whereas some others adversely affect the partition.

The application of the methodology to the analysis of playing styles shows decidedly interesting results: the building up of the offensive manoeuvre from the lateral 
zones of the field has a positive effect on the number of goals scored by a team. The high internal heterogeneity of groups - necessary to obtain a manageable 
number of clusters - obviously lessens the effectiveness of the approach, and consequently the effect of the  tactics detected on the final outcome of a match.
 
Unfortunately, the adjustment made to the Dixon and Coles model does not allow  the final result of a game to be forecast as the teams' playing styles are available only at 
the end of the match. To use the proposed method in this context, we plan to construct a model to predict the tactic of a team on the basis of its previous performances;
additionally, thanks to the proliferation of in-running betting, it is also possible to use the information about the playing style during the first half of the match
to predict the overall tactic in a certain game.



\section*{Supplementary material}

\subsection*{Appendix A}

The following analysis focuses on the features that may mark a generic population of networks (see Section \ref{sectionsimilaritybetweennetworks}).

\begin{figure}
\begin{center}
\includegraphics[width=\textwidth,height=7.5cm]{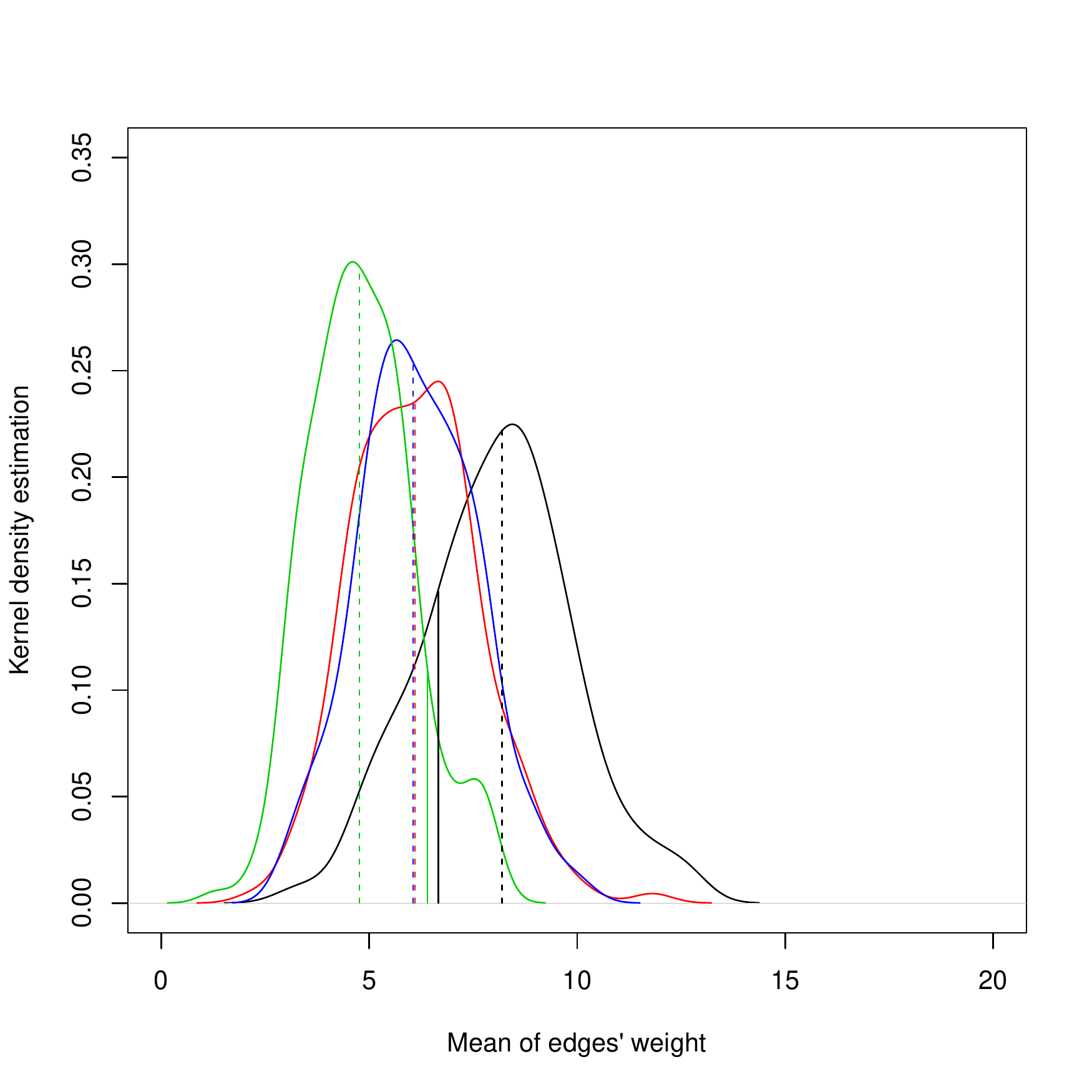} 
\caption{Kernel density estimation of the average weight of the edges for high ranking teams (black line), middle-high ranking teams (red line), middle-low ranking teams (blue line) and low ranking teams (green line). The dashed lines display the medians, while the continuous lines display the first quintile (black line) and the ninth decile (green line) of the respective distributions.}
\label{PesomediodegliarchiNUOVO}
\end{center}
\end{figure}

The first aspect to investigate is the average weight of the different units; since the Italian teams differ considerably as regards passing ability, one is justified in expecting this value to vary according to the network considered. 
Figure \ref{PesomediodegliarchiNUOVO}
shows the kernel density estimation of the average weight according to the team's position in the standings at the end of the season. Specifically, the teams are divided into four categories, high ranking (positions 1-5), middle-high ranking (positions 6-10), middle-low ranking (positions 11-15) and low ranking (positions 16-20). The figure seems to suggest that teams with middle-high and middle-low rankings have similar distributions, whereas the top teams differ greatly  from the teams in the relegation zone. Indeed, the first quintile of the rightmost distribution is greater than the ninth decile of the leftmost one.

Another element of particular interest concerns the connections between the areas of the pitch: also in this case there is high heterogeneity, with an average value ranging from 0.11 - observable in the relationship between zone 3 (left defense) and  zone 7 (right attack), see Figure \ref{IMMsuddivisionecampodacalcio}  - to 25.59, observable in the relationship between zone 4 (left midfield) and zone 5 (central midfield). In particular, the two edges linking nodes (3,7) and (1,9) - left defense/right attack and right defense/left attack respectively - present a highly asymmetrical distribution of weights: in addition to the inflation of zero (91\% and 89\% respectively), they are characterised by the minimum variance (0.12 and 0.15 respectively) and the lowest maximum value, i.e. three. This evidence is not completely surprising as the nodes involved refer to areas diametrically opposed on the pitch: the almost complete absence of connections suggests that the ones which were detected are accidental and fully assimilable to the long passes which ended up off the pitch (and so were not counted). As a consequence, in the application of the clustering method all the weights of the abovementioned edges are set to zero in order not to modify the community structures on the basis of fortuitous events.

The graphical-descriptive analysis can be also informative with regard to a question of vital importance concerning the utility of the proposed methodology: the approach described in Section \ref{sectionproposedmethodology} requires a preprocessing phase aimed at managing the characteristics of the sample which, if it is of little use, represents a meaningless complication compared to a simpler comparison between the community structures of the starting networks. Let's consider two contrasting teams as regards the quality of ball possession: Napoli and Frosinone. The former showed incredibly dense possession as evidenced by the maximum value for the median, third quartile and maximum concerning the average number of connections between nodes during a match. On the other hand the latter, who were dramatically relegated to Serie B ConTe.it at the end of the season, showed sporadic possession as evidenced by the minimum value for the first quartile, mean, median and third quartile concerning the same quantity. For each of the 38 matches played by each of the two teams, it is possible to calculate the ratio of a specific edge's weight 
to the overall weight of the network in order to find differences between the two teams without considering the overall number of connections. By comparing the average values of these quantities between the two teams, it is noticeable that 10 of the 11 edges with the lowest relative weight - accounting for 1.52\% and 3.57\% respectively of Napoli's and Frosinone's overall weight - are common to the two teams, as are 10 of the 11 connections with the highest relative weight - accounting for 74.28\% and 70.26\% respectively -. Table \ref{zonepiumenocollegate} summarises this evidence.
In view of the above, there is a clear consideration to be made: the directions in which the playing styles are mainly developed are shared by the teams due to the football-specific dynamics, i.e. in statistical terms due to the characteristics of the population studied. As a consequence, a clustering method considering the broader framework in which the single network is placed is strongly recommended.
\begin{table} []
\centering 
\begin{tabular}{r r r | r r r}
\hline
$\downarrow$ Edge & \% Napoli & \% Frosinone & $\uparrow$ Edge & \% Napoli & \% Frosinone\\ 
\hline
(4,1) & -  &0.404 &                               (5,4)&14.78&7.48\\
(9,6) & 0.410 &0.373&                         (6,5)&10.77&6.81\\
(9,3) & 0.207 &0.446&                         (9,4)&9.53&7.05\\
(6,3) & 0.207&0.287&                          (7,6)&6.31&7.28\\
(9,2) & 0.191 &0.454&                         (4,3)&6.13&6.79\\
(8,2) & 0.122 &-&                                 (3,2)&5.15&6.96\\
(7,2) &  0.098 &0.455&                        (6,1)&4.79&5.97\\
(7,1) & 0.082&0.365&                          (9,8)&4.65&-\\
(8,3) & 0.074 & 0.201&                        (5,2)&4.37&6.97\\
(9,1) & 0.051 &0.157&                         (2,1)&3.97&5.85\\
(8,1) &  0.039 &0.267&                        (8,7)&3.83&4.47\\
(7,3) & 0.036 & 0.160 &                       (6,2)&-&4.63\\
\hline
Total & 1.52 &3.57&Total&74.28&70.26\\

\end{tabular}
\caption{Labels and relative weights of the edges with the lowest and highest relative weight for Napoli and Frosinone. }
\label{zonepiumenocollegate}
\end{table}

\subsection*{Appendix B}
The following brief simulation study is aimed at evaluating the suitability of the proposed clustering method. In particular, the impact of the value chosen for
$\alpha$ on the reliability of the methodology is analyzed. Moreover, the results are compared to those obtained by an alternative approach. 

\textbf{Scenarios}

Let $R=\{ R_1,\dots,R_n \}$ be a sample of $n$ undirected weighted networks, with $\mathcal{K}=\{k_1, k_2,\dots,k_K\}$ the set of nodes shared between the networks. 
In addition, let the sample be partitioned into two groups, with $n_1$ and $n_2$ the respective sizes. In all the scenarios, we set $n=100,  K=30,  n_1=n_2=50$. For the sake
of simplicity, the nodes are labelled with numbers from 1 to 30 and with an even/odd node we refer to a node labelled with an even/odd number, respectively. 

\begin{figure}[]
\centering%
{\includegraphics[width=0.75\textwidth,height=6cm]{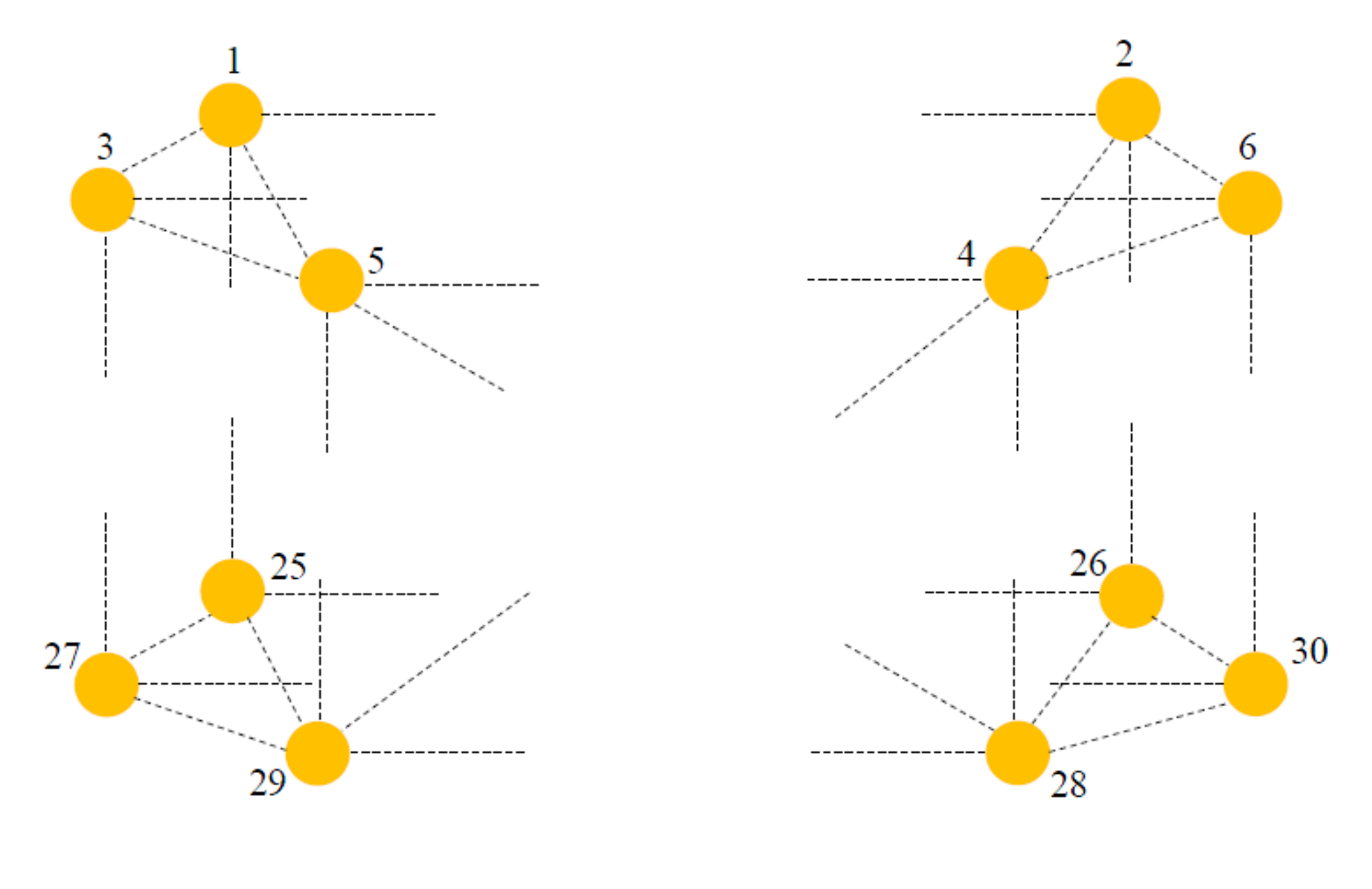}}\qquad \qquad
{\includegraphics[width=0.75\textwidth,height=6cm]{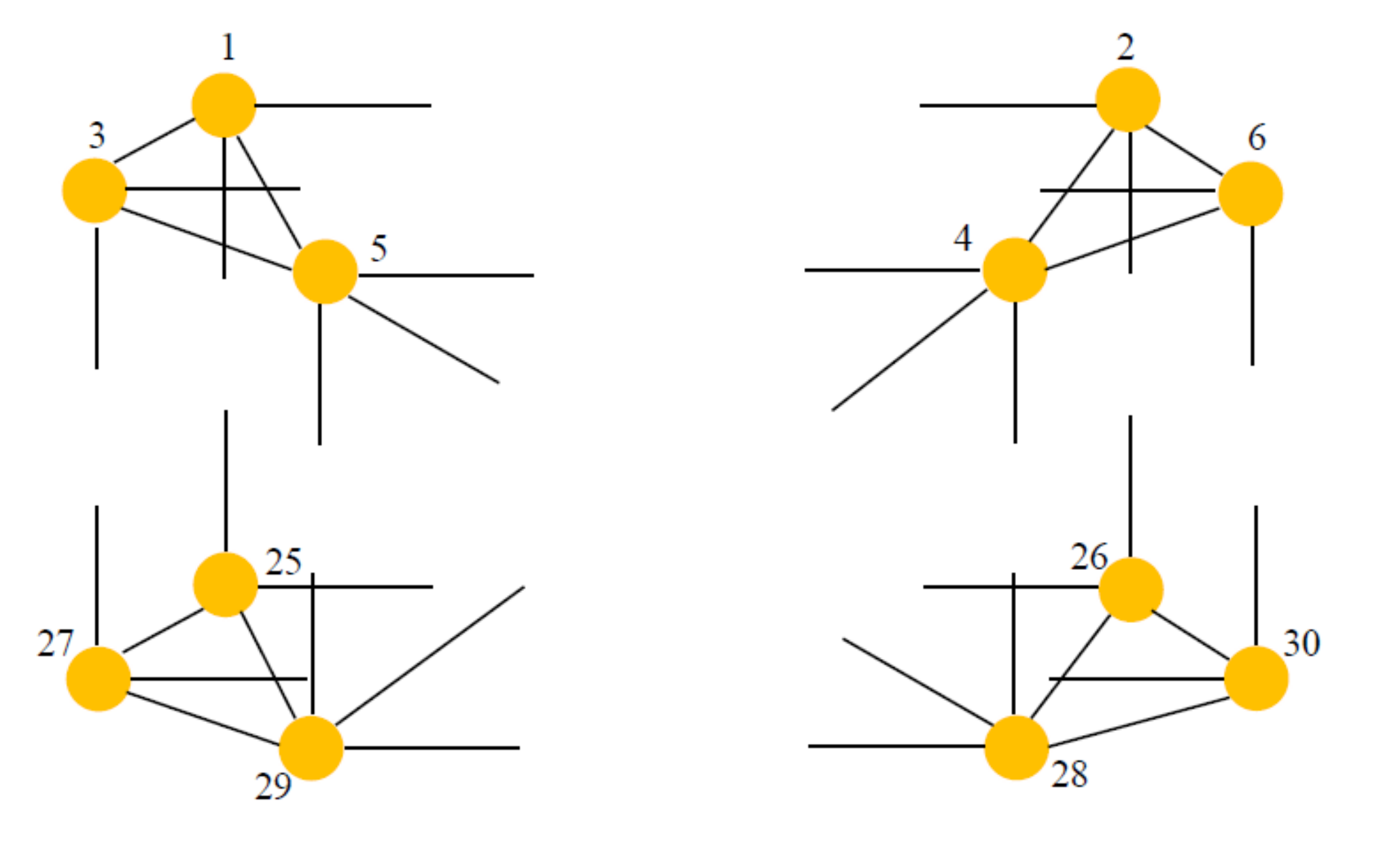}} 
\caption{Scenario 1. A network belonging to the first group (at the top) and a network belonging to the second group (at the bottom) are represented: the dashed lines represent connections generated independently from $X$, the continuous lines represent connections generated independently from $Y$. }
\label{networkscenario1}

\end{figure}

\begin{figure}[]
\centering%
{\includegraphics[width=0.75\textwidth,height=6cm]{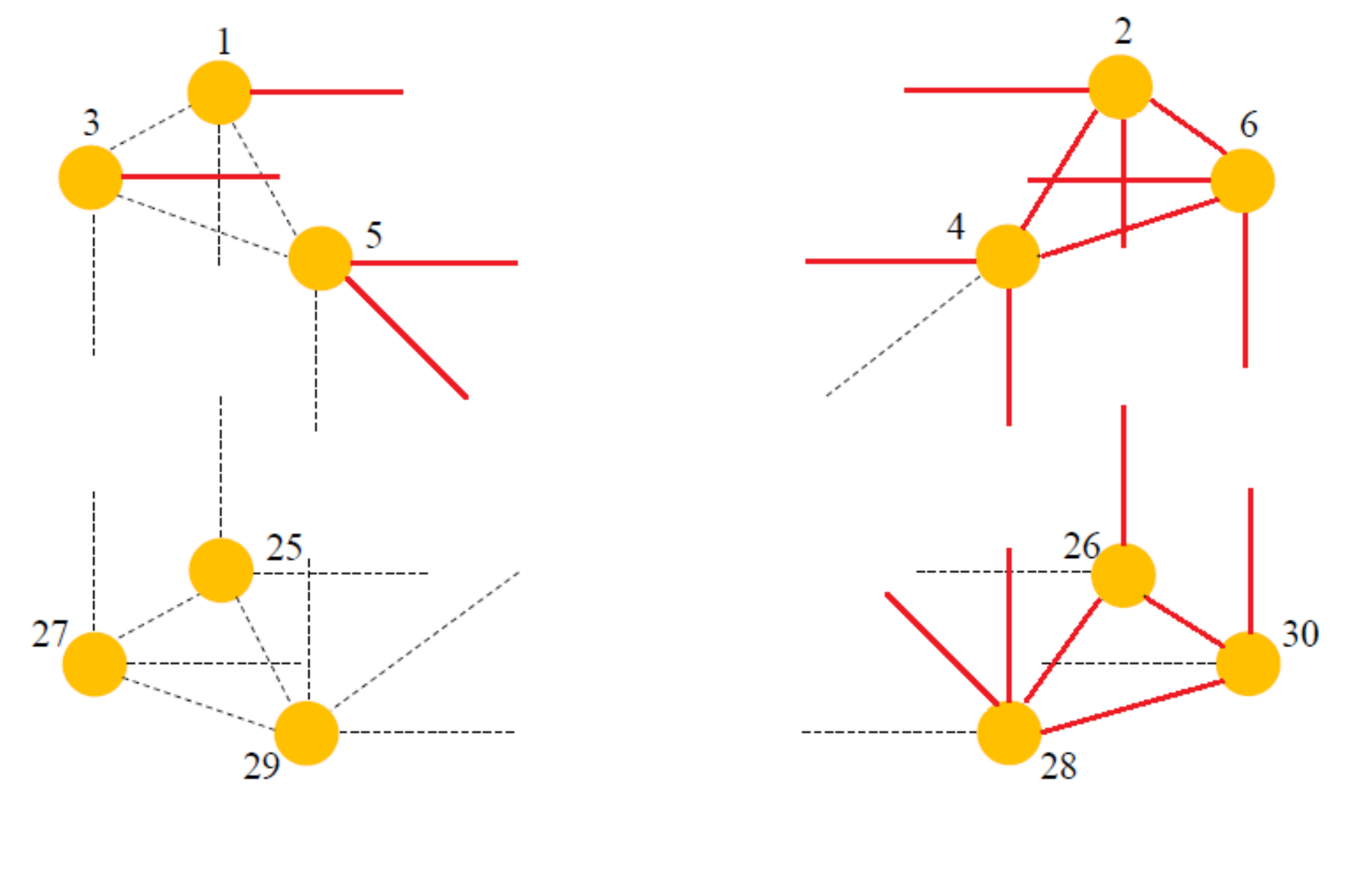}}\qquad \qquad
{\includegraphics[width=0.75\textwidth,height=6cm]{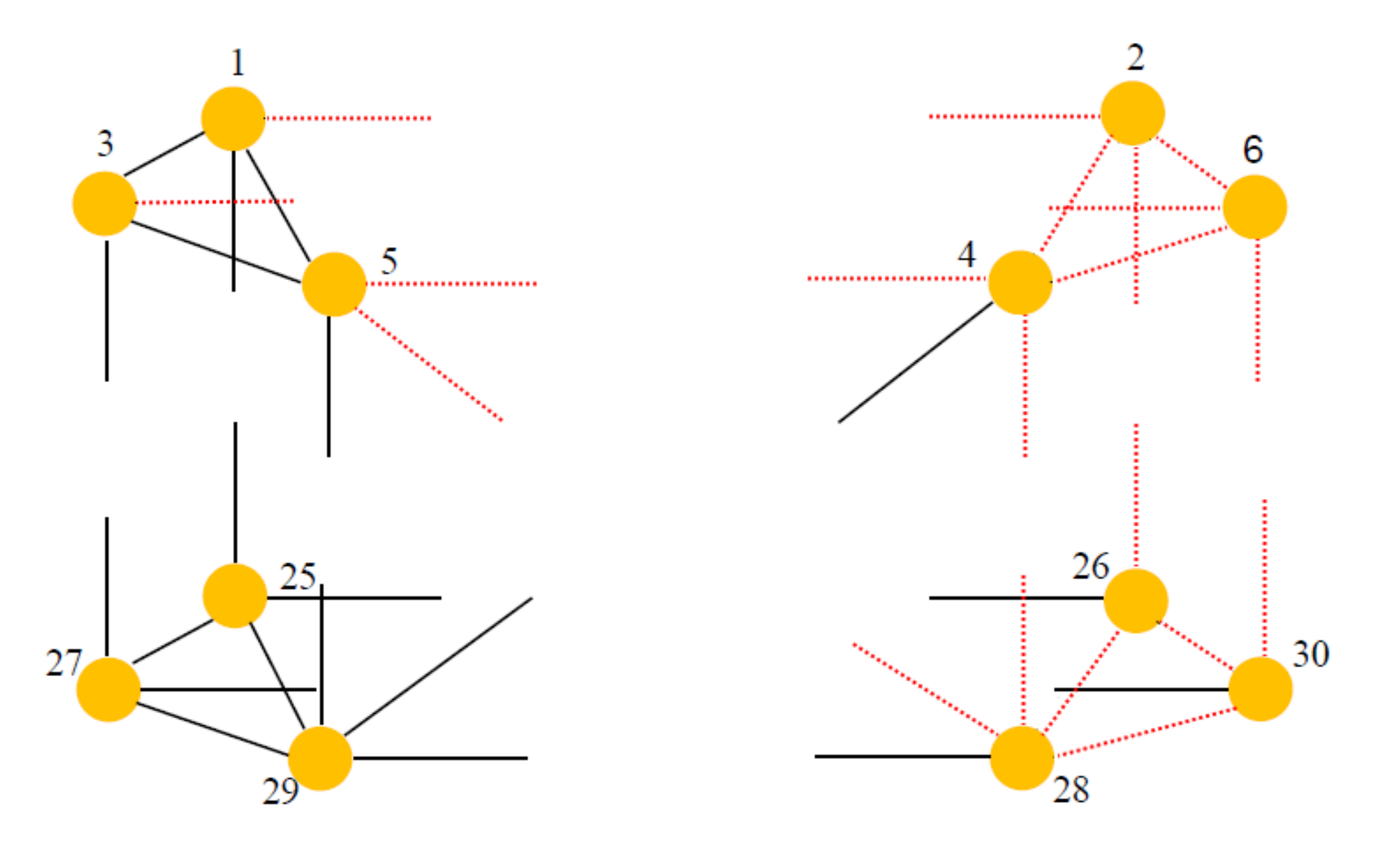}} 
\caption{Scenario 2. A network belonging to the first group (at the top) and a network belonging to the second group (at the bottom) are represented: the continuous red lines represent connections generated independently from $X_1$, the dashed black lines represent connections generated independently from $X_2$, the dashed red lines represent connections generated independently from $Y_1$, the continuous black lines represent connections generated independently from $Y_2$.}
\label{fanculo}
\end{figure}

The study is structured into two scenarios, subdivided in their turn  into two subscenarios of differing complexity. A graphical representation of the first scenario is provided in Figure \ref{networkscenario1}, of the second one in Figure \ref{fanculo}. 

In the first scenario, the networks belonging to the first group are characterised by connections between nodes generated independently from a discrete random variable $X$,
whereas those belonging to the second group are characterised by connections between nodes generated independently from a discrete random variable $Y$. 
Thus:
\begin{itemize}[noitemsep,topsep=0pt,parsep=0pt,partopsep=0pt]
\item \textit{First subscenario}: $D_X=\{4,5,6,7\}$ the support of $X$,  $D_Y=\{10,11,12,13\}$ the support of $Y$ and $\pi_X=\pi_Y=\{0.4,0.3,0.2,0.1\}$ the probability vector of $X$ and $Y$.
\item \textit{Second subscenario}: $D_X=\{4,5,6,7,8\}$ the support of $X$, $D_Y=\{5,6,7,8,9\}$ the support of $Y$ and $\pi_X=\pi_Y=\{0.1,0.25,0.3,0.25,0.1\}$ the probability vector of $X$ and $Y$.
\end{itemize}

The community structures are completely random for all the units, but the networks of the first group differ from those of the second group with 
regard to their overall weight. The second subscenario displays a higher complexity than the first one since the difference in the overall weight between the two 
groups drops.
The purpose of this scenario is to replicate one of the features shown in Appendix A, namely
that the average weight of the different units varies according to the team considered.

In the second scenario:
\begin{itemize}[noitemsep,topsep=0pt,parsep=0pt,partopsep=0pt]
\item the weight of each edge linking two \emph{even} nodes is generated independently from the discrete random variable $X_1$ in the first group and from the discrete 
random variable $Y_1$ in the second group.
\item the weight of each edge linking two \emph{odd} nodes is generated independently from the discrete random variable $X_2$ in the first group and from the discrete 
random variable $Y_2$ in the second group.
\item the weight of each edge linking an \emph{even} node with one of the first eight \emph{odd} nodes \{1,3,\dots,13,15\} is generated independently from $X_1$ in the first group and from $Y_1$ in the second group.
\item the weight of each edge linking an \emph{even} node with one of the remaining \emph{odd} nodes \{17,19,\dots,27,29\} is generated independently from $X_2$ in the first group and from $Y_2$ in the second group.
\end{itemize}
For the sake of simplicity, let us define $E_{X_1}$ ($E_{X_2},E_{Y_1},E_{Y_2}$ respectively) the set of edges whose weights are generated from $X_1$ ($X_2, Y_1, Y_2$ respectively).  
Thus:
\begin{itemize}[noitemsep,topsep=0pt,parsep=0pt,partopsep=0pt]
\item \textit{First subscenario}: $D_{X_1}=\{10,11,12\}, D_{X_2}=\{2,3,4\}, D_{Y_1}=\{9,10,11\}, D_{Y_2}=\{3,4,5\}$ the supports of $X_1, X_2, Y_1, Y_2$;  $\pi_{X_1}=\{0.46,0.46,0.08\}, \\ \pi_{X_2}=\{0.8,0.1,0.1\}, 
\pi_{Y_1}=\{0.64,0.18,0.18\}, \pi_{Y_2}=\{0.1,0.1,0.8\}$ the probability vectors of $X_1, X_2, Y_1,Y_2$. 
\item \textit{Second subscenario}: $D_{X_1}=\{12,13,14,15,16,17\}, D_{X_2}=\{1,2,3,4,5,6\}, D_{Y_1}=\{11,12,13,14,15,16\}, D_{Y_2}=\{2,3,4,5,6,7\}$ the supports of $X_1, X_2, Y_1, Y_2$;  \\ $\pi_{X_1}=\pi_{Y_2}=\{0.18,0.18,0.18,0.18,0.18,0.10\}, \pi_{X_2}=\pi_{Y_1}= \\ \{0.10,0.18,0.18,0.18,0.18,0.18\}$  the probability vectors of $X_1, X_2, Y_1,Y_2$. 
\end{itemize}

The second subscenario displays a higher complexity than the first one since the difference in the weights of the edges between the two groups drops; the purpose of this scenario is to replicate another one of the features shown in Appendix A, namely the one where the teams share the directions in which the playing styles are mainly developed but where some different trends between the teams can be detected in the densely and sparsely connected areas of the pitch .

Every scenario is developed for three values of $\alpha$: 0.3, 0.7, 0.95.
The methodology is compared to the \emph{naive} approach introduced in Section \ref{sectionsimilaritybetweennetworks}: it consists of considering each starting network on its own, detecting its community 
structure and merging, step by step, the networks whose partitions are most similar. Even in this case every self-loop is set to zero for the reason described in 
Section \ref{paragrafofase2}.
The suitability of the methods is examined by comparing the partition obtained requiring $N_G=2$ with the real subdivision of the sample; to do so we use the ARI \citep{hubert1985comparing}. The simulations are achieved by using programming language R \citep{Rsoftware}, the detection of community structures by 
\emph{igraph} package \citep{igraph} and the computation of ARI by \emph{pdfCluster} package \citep{pdfCluster}. Every scenario is evaluated using $N=1000$ replications.

\textbf{Results}

\begin{figure}[]
\centering%
{\includegraphics[width=0.75\textwidth,height=11cm]{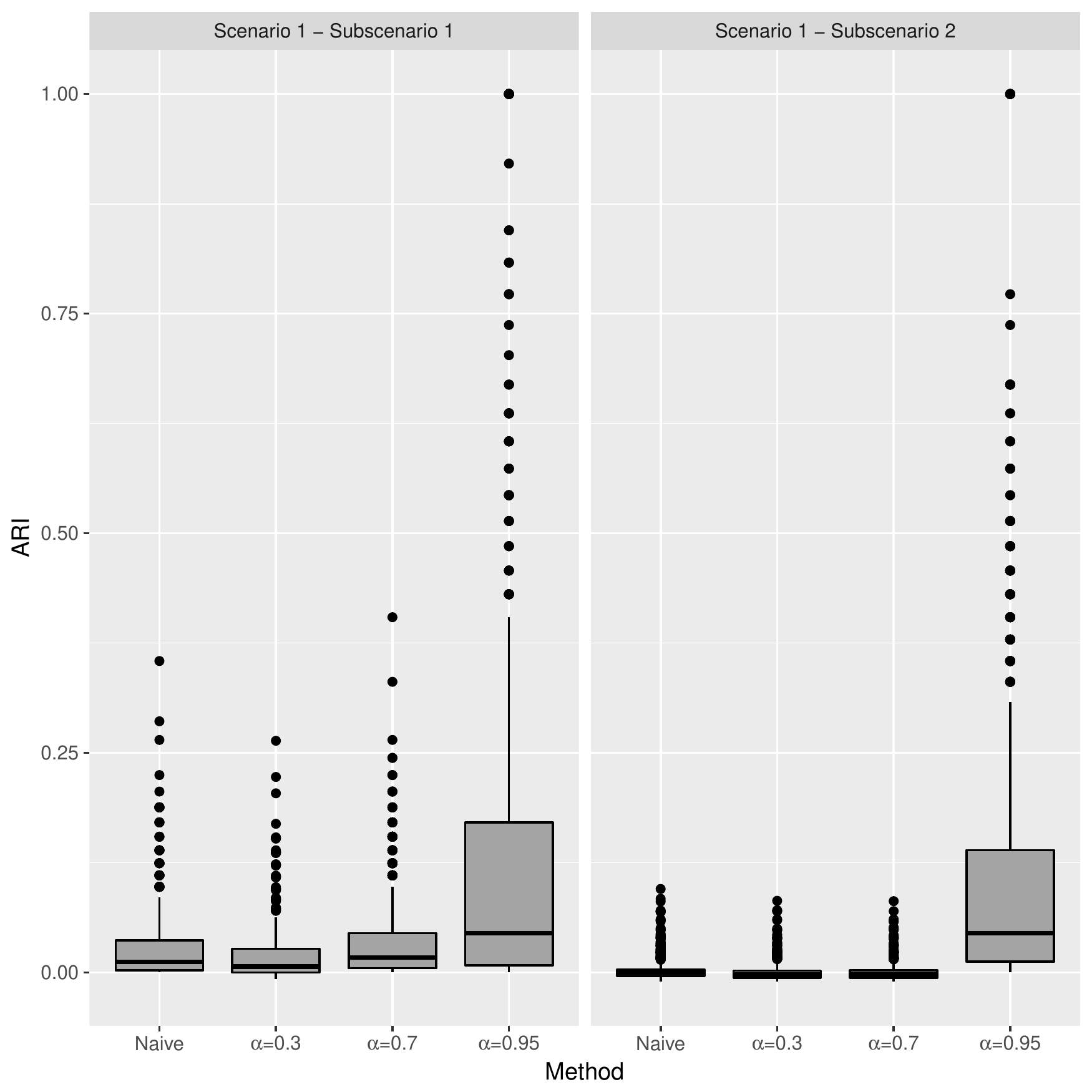}}\qquad \qquad
{\includegraphics[width=0.75\textwidth,height=11cm]{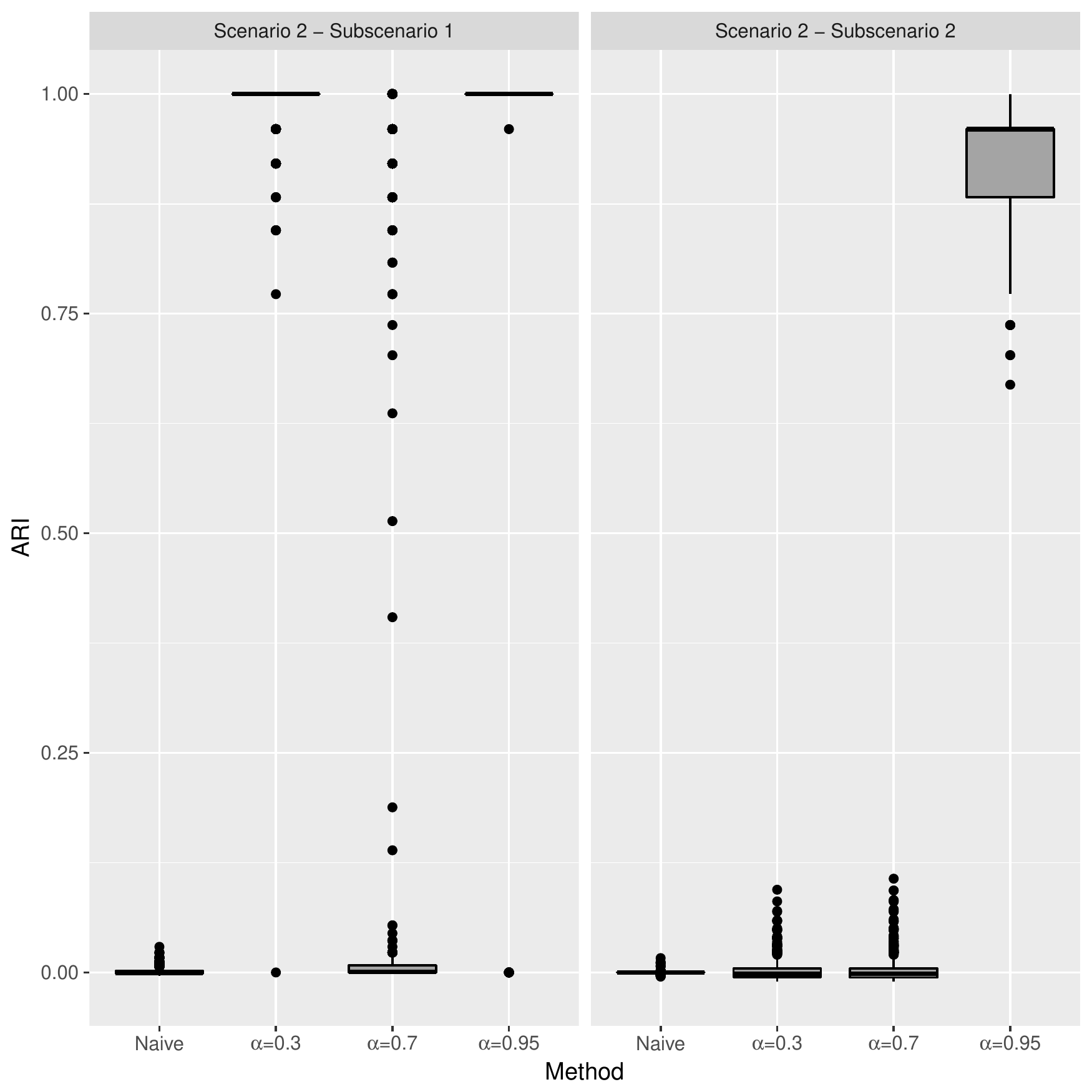}} 
\caption{Results of the simulation study.}
\label{figurasimulazioni}
\end{figure}

Figure \ref{figurasimulazioni} displays the box-plots of ARI concerning the 1000 replications for each scenario.
The first scenario shows an improvement in the method when the highest value for $\alpha$ is considered: the choice of $\alpha=0.95$ allows only the particularly 
surprising connections for the population to be included and therefore enables the differences between the two groups to be detected. The low values of the agreement with
the real partition are easily explained since the arbitrary nature of community structures involves networks belonging to the same group having, by construction, low similarity.
A decidedly positive aspect is noticed when moving through subscenarios: the performance of the 
procedure remains almost unchanged despite the fact that the groups are less dissimilar.

The second scenario shows a complete failure in the \emph{naive} approach: 
the fact that $E_{X_1}=E_{Y_1}$ and $E_{X_2}=E_{Y_2}$
between groups involves the community structures between the two subpopulations being identical, and so the normalization of the weights seems to be a fundamental adjustment to improve the quality of the partitions.
The remarks concerning the role of $\alpha$ vary according to the subscenario considered: in the first one the choices of $\alpha=0.30$ and $\alpha=0.95$ allow the differences between the two groups to be managed, whereas intermediate value $\alpha=0.70$ produces decidedly unsatisfactory results. On the other hand, in the second one the differences between groups  are so minimal that only a value as high as $\alpha=0.95$ achieves the goal.

Overall, the methodology provides a reasonable compromise in the handling of the different aspects characterising the clustering of networks, but the impact of the
value of $\alpha$ on the results represents an undeniable limit in an unsupervised context.

\subsection*{Appendix C}
\begin{figure}
\begin{center}
\includegraphics[width=\textwidth,height=18cm]{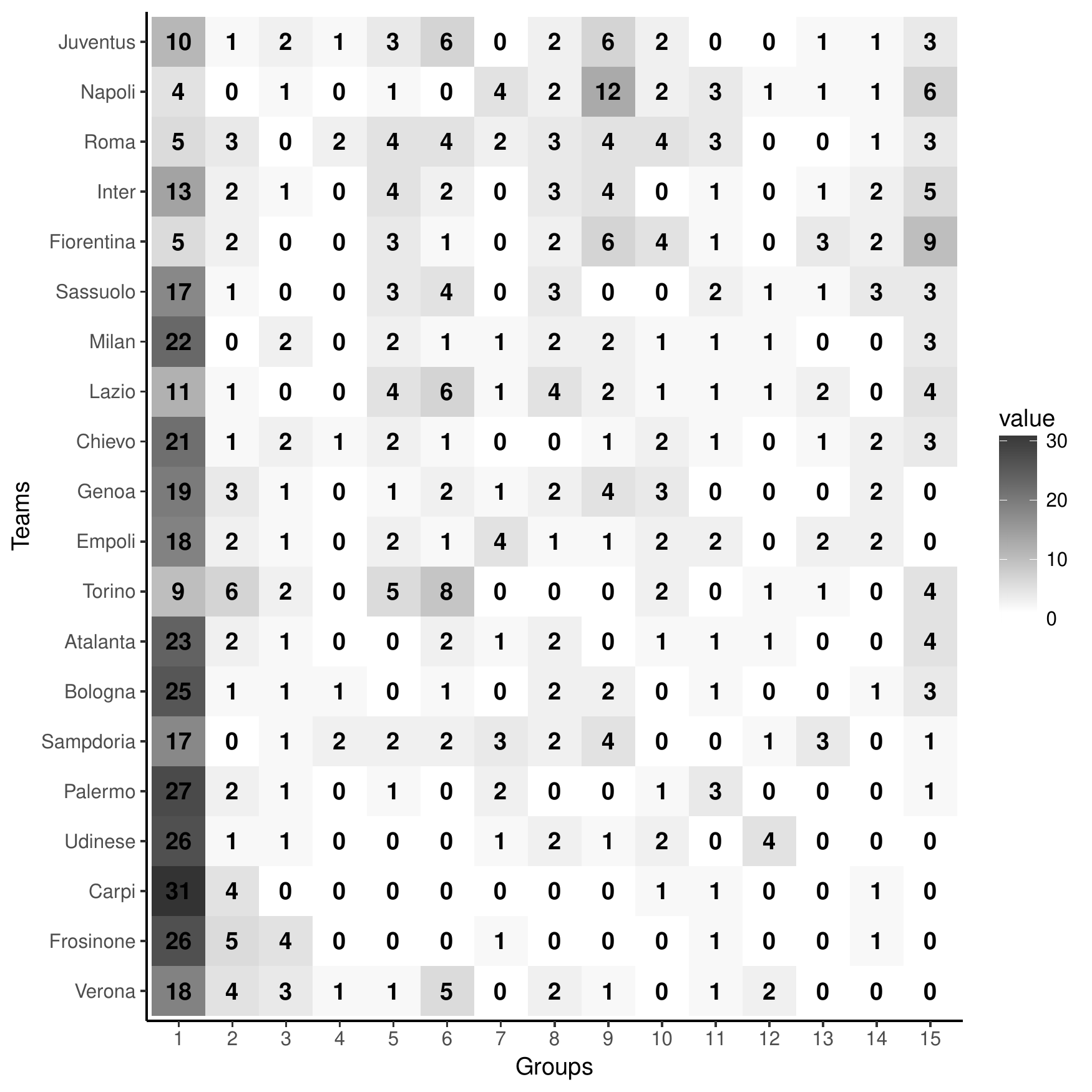} 
\caption{Number of networks allocated, for each team, to each group.}
\label{tabellagruppisquadre}
\end{center}
\end{figure}

The objective criterion used to identify the main characteristics of a specific group consists of the evaluation of the percentage of times that, considering the networks of that cluster, pairs and triads
of nodes are allocated to the same community: a high value for this quantity means that the group is made up of playing styles with frequent connections between those 
pairs and triads of areas. 
A brief description of the groups is provided below (for all the two-dimensional and three-dimensional arrays reporting the described percentages, contact J.Diquigiovanni). For the sake of clarity, the clusters are divided into 6 categories identifying the main macro typologies of
playing styles. Figure \ref{tabellagruppisquadre} shows the number of networks allocated to each group for each team.

\textbf{Sparse playing style}
\begin{itemize}[noitemsep,topsep=0pt,parsep=0pt,partopsep=0pt]
\item \textit{Group 1}, made up of 347 networks. The group includes almost half of the units, which are characterised by the absence of any significant connection between
the areas of the pitch; the size of the cluster is not entirely surprising as the value chosen for $\alpha$ is particularly high. It is mainly made up of low ranking
teams, which are allocated to this group 67\% of the time; on the other hand, Napoli (four out of 38 networks), Roma (five out of 38 networks) and
Fiorentina (five out of 38 networks) are the least involved teams.
\end{itemize}
\textbf{Long ball playing style}
\begin{itemize}[noitemsep,topsep=0pt,parsep=0pt,partopsep=0pt]
\item \textit{Group 2}, made up of 41 networks. This playing style is characterised by long passes between the right defense zone and the central attack zone (1-8). 
Despite the fact that the group is mainly made up of middle-low and low ranking teams (66\%), there is no lack of high ranking teams (20\%).

\item \textit{Group 3}, made up of 24 networks. This playing style is characterised by long passes involving the central defense zone (2-8, 2-9). The low ranking teams are the most 
numerous in the group (38\%).

\item \textit{Group 4}, made up of 8 networks. This playing style is characterised by long passes between the central defense zone and the right attack zone (2-7): all the networks
of the group share this community. The use of this tactic is  distributed homogeneously between the teams.
\end{itemize}
\textbf{Dense playing style}
\begin{itemize}[noitemsep,topsep=0pt,parsep=0pt,partopsep=0pt]
\item \textit{Group 5}, made up of 38 networks. This playing style is characterised by ball possession in the defense zone (1-3, 1-2-3). The high and middle-high ranking teams are 
the most numerous in the group (71\%).

\item \textit{Group 6}, made up of 46 networks. This playing style is characterised by connections between the central defense zone and the lateral midfield zones (2-4, 2-6, 2-4-6).
The low ranking teams are the least numerous in the group (11\%).

\item \textit{Group 7}, made up of 21 networks. This playing style is characterised by connections between the central defense zone and the central/right midfield zones (2-5, 5-6,
2-5-6). The use of this tactic is  distributed homogeneously between the teams.

\item \textit{Group 8}, made up of 34 networks. This playing style is characterised by ball possession in the offensive zones (4-7, 8-9). The high and middle-high ranking teams are 
the most numerous in the group (68\%).

\item \textit{Group 9}, made up of 50 networks. This playing style is characterised by ball possession in the offensive zones (4-9,5-7,6-7,4-8-9,5-6-7). The high ranking teams are 
the most numerous in the group (64\%).
\end{itemize}
\textbf{Mixed playing style}
\begin{itemize}[noitemsep,topsep=0pt,parsep=0pt,partopsep=0pt]
\item \textit{Group 10}, made up of 28 networks. 
This playing style is characterised by long passes between the left defense zone and the left attack zone (3-9) and by connections between the left defense zone and the central midfield zone (3-5). 
The high and middle-high ranking teams are the most 
numerous in the group (68\%).

\item \textit{Group 11}, made up of 23 networks. 
This playing style is characterised by long passes between the left defense zone and the central attack zone (3-8) and by connections between the right defense zone and the central midfield zone (1-5). 
The use of this tactic is  distributed homogeneously between the teams.
\end{itemize}
\textbf{Rapid attack playing style}
\begin{itemize}[noitemsep,topsep=0pt,parsep=0pt,partopsep=0pt]
\item \textit{Group 12}, made up of 13 networks. 
This playing style is characterised by connections between the right midfield zone and the central attack zone (6-8). 
The low ranking teams are the most numerous in the group (46\%).

\item \textit{Group 13}, made up of 16 networks. 
This playing style is characterised by connections between the central midfield zone and the central attack zone (5-8). 
The low ranking teams are not present in this group.

\item \textit{Group 14}, made up of 19 networks. 
This playing style is characterised by connections between the left midfield zone and the central attack zone (4-8) and between the lateral zones of the attack (7-9).
The high and middle-high ranking teams are the most numerous in this group (74\%).
\end{itemize}
\textbf{On-the-wings playing style}
\begin{itemize}[noitemsep,topsep=0pt,parsep=0pt,partopsep=0pt]
\item \textit{Group 15}, made up of 52 networks. 
This playing style is characterised by connections between the lateral zones of the defense and the adjacent lateral zones of the midfield (1-6,3-4): 87\% of the networks present at least one of the two communities. The high ranking teams are the most numerous (50\%), with Fiorentina (nine networks) and Napoli (six networks) the most involved teams. In view of the importance of this cluster (see Section \ref{modellingfinalscores}), a graphical representation is provided in Figure \ref{gruppo13}.
\begin{figure}
\begin{center}
\includegraphics[width=0.75\textwidth,height=7cm]{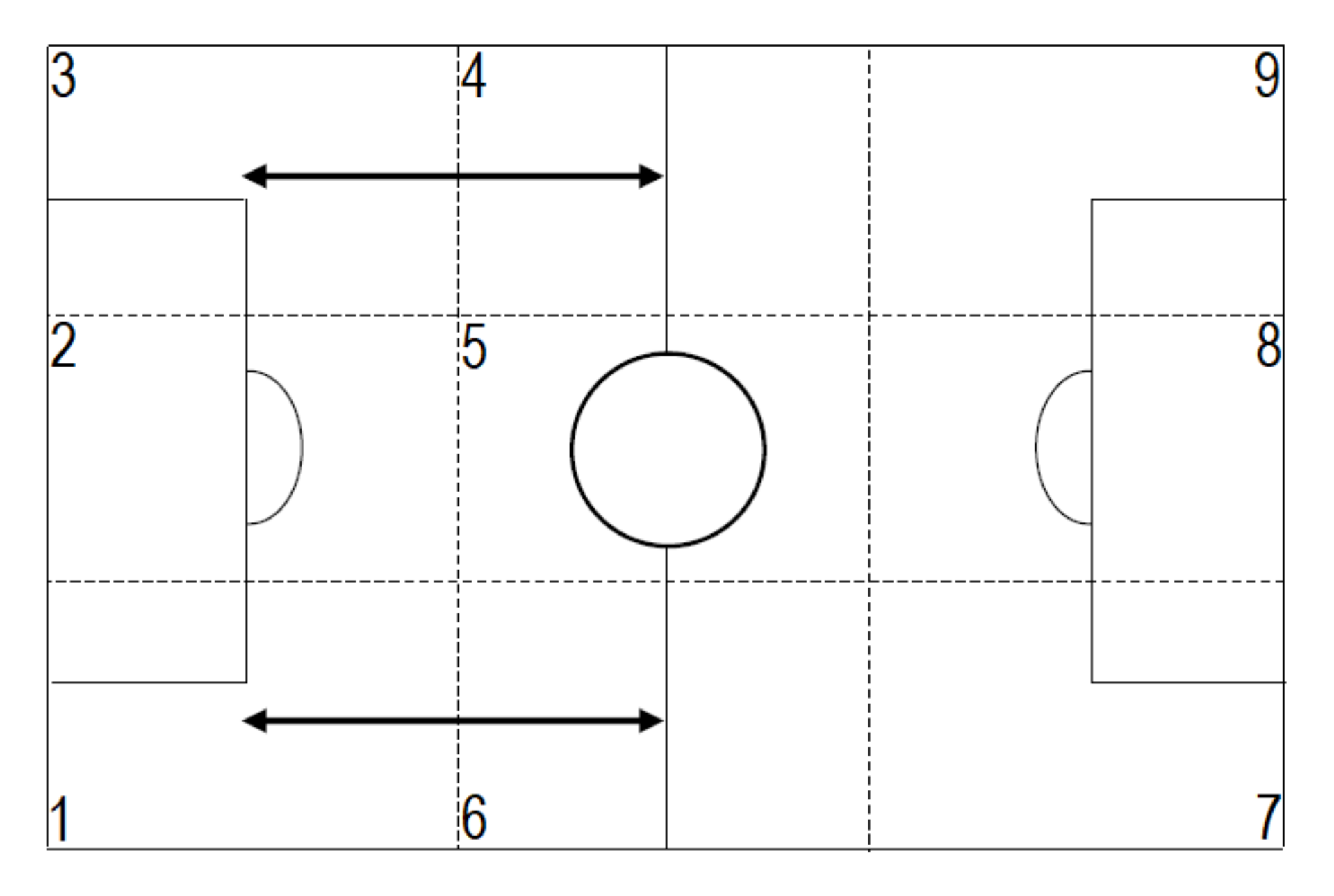} 
\caption{Graphical representation of the on-the-wings playing style.}
\label{gruppo13}
\end{center}
\end{figure}
\end{itemize}

In view of the above, it is now possible to extrapolate some information about either the league or the teams. For example, Juventus were on an incredible winning streak (26 out of 28 matches) after a disastrous start of the season characterised by only 12 points in the first ten matches. Consequently, it is certainly surprising that only two of the ten matches in which Juventus displayed a sparse playing style (Group 1) refer to that critical starting period. This evidence, in spite of its simplicity, seems to confirm a clich\'e on Italian football: the team which plays better does not always win.

\section*{Acknowledgements}
This work was partially funded by grant CPDA154381/15 from the University of Padova, Italy. The authors are grateful to Lazar Petrov, Italian lead of InStat, and Lorenzo Favaro, CEO of SportAnalisi, who provided data. We also thank Daniele Durante and Nicola Sartori for fundamental comments and advices, and David Dandolo for providing the code on the Dixon and Coles model \citep{dandolo2017modellazione}.


\begin{thebibliography}{99}

\bibitem[Anderson and Sally (2013)]{anderson2013numbers}
Anderson, C. and Sally, D. (2013). \textit{The numbers game: Why everything you know
about soccer is wrong}. Penguin.

\bibitem[Azzalini and Menardi (2014)]{pdfCluster}
Azzalini, A. and Menardi, G. (2014). Clustering via nonparametric density estimation:
The R package pdfCluster. \textit{Journal of Statistical Software}, \textbf{57}(11), 1–26. URL \url{http://www.jstatsoft.org/v57/i11/}.


\bibitem[Azzalini and Scarpa (2012)]{azzalini2012data}
Azzalini, A. and Scarpa, B. (2012). \textit{Data analysis and data mining: An introduction}.
Oxford University Press USA.

\bibitem[Baio and Blangiardo (2010)]{baio2010bayesian}
Baio, G. and Blangiardo, M. (2010). Bayesian hierarchical model for the prediction of football results. \textit{Journal of Applied Statistics}, \textbf{37}(2), 253–264.


\bibitem[Blondel et al. (2008)]{blondel2008fast}
Blondel, V. D., Guillaume, J.-L., Lambiotte, R., and Lefebvre, E. (2008). Fast unfolding of communities in large networks. \textit{Journal of statistical mechanics: theory
and experiment}, \textbf{2008}(10), P10008.

\bibitem[Borrie et al. (2002)]{borrie2002temporal}
Borrie, A., Gudberg K. J., and Magnus S. M. (2002). Temporal pattern analysis and its applicability in sport: an explanation and exemplar data. \textit{Journal of sports sciences} \textbf{20}.10: 845-852.

\bibitem[Brandes et al. (2006)]{brandes2006maximizing}
Brandes, U., Delling, D., Gaertler, M., G{\"o}rke, R., Hoefer, M., Nikoloski, Z., and Wagner, D. (2006). Maximizing modularity is hard. \textit{arXiv preprint physics/0608255}.

\bibitem[Brandt and Brefeld (2015)]{brandt2015graph}
Brandt, M. and Brefeld, U. (2015). Graph-based approaches for analyzing team interaction on the example of soccer. In \textit{Proceedings of the ECML/PKDD Workshop on Machine Learning and Data Mining for Sports Analytics}.

\bibitem[Cintia et al. (2015a)]{cintia2015harsh}
Cintia, P., Giannotti, F., Pappalardo, L., Pedreschi, D., and Malvaldi, M. (2015a).
The harsh rule of the goals: Data-driven performance indicators for football teams.
In \textit{2015 IEEE International Conference on Data Science and Advanced Analytics
(DSAA) , Paris, oct 2015}, pages 1–10. IEEE.

\bibitem[Cintia et al. (2015b)]{cintia2015network}
Cintia, P., Rinzivillo, S., and Pappalardo, L. (2015b). A network-based approach to
evaluate the performance of football teams. In \textit{Proceedings of the ECML/PKDD
Workshop on Machine Learning and Data Mining for Sports Analytics}.


\bibitem[Clemente et al. (2015a)]{clemente2015using}
Clemente, F. M., Couceiro, M. S., Martins, F. M. L., and Mendes, R. S. (2015a).
Using network metrics in soccer: a macro-analysis. \textit{Journal of human kinetics}, \textbf{45}(1), 123–134.

\bibitem[Clemente et al. (2015b)]{clemente2015social}
Clemente, F. M., Martins, F. M. L., Kalamaras, D., Oliveira, J., Oliveira, P., and Mendes, R. S. (2015b). The social network analysis of switzerland football team
on fifa world cup 2014. \textit{Journal of Physical Education and Sport}, \textbf{15}(1), 136.

\bibitem[Csardi and Nepusz (2006)]{igraph}
Csardi, G. and Nepusz, T. (2006). The igraph software package for complex network
research. \textit{InterJournal}, \textbf{Complex Systems}, 1695. URL \url{http://igraph.org}.

\bibitem[Dandolo (2017)]{dandolo2017modellazione}
Dandolo, D. (2017). Modellazione statistica di risultati calcistici. \textit{Bachelor’s thesis}.

\bibitem[Dixon and Coles (1997)]{dixon1997modelling}
Dixon, M. J. and Coles, S. G. (1997). Modelling association football scores and
inefficiencies in the football betting market. \textit{Journal of the Royal Statistical Society:
Series C (Applied Statistics)}, \textbf{46}(2), 265–280.



\bibitem[Durante et al. (2017) ]{durante2016nonparametric}
Durante, D., Dunson, D. B., and Vogelstein, J. T. (2017). Nonparametric bayes modeling of populations of networks. \textit{Journal of the American Statistical Association},
pages 1–15.

\bibitem[Geys et al. (1999)]{geys1999pseudolikelihood}
Geys, H., Molenberghs, G., and Ryan, L. M. (1999). Pseudolikelihood modeling
of multivariate outcomes in developmental toxicology. \textit{Journal of the American
Statistical Association}, \textbf{94}(447), 734–745.



\bibitem[Godambe (1960)]{godambe1960optimum}
Godambe, V. P. (1960). An optimum property of regular maximum likelihood esti
mation. \textit{The Annals of Mathematical Statistics}, \textbf{31}(4), 1208–1211.


\bibitem[Gyarmati et al. (2014)]{gyarmati2014searching}
Gyarmati, L., Kwak, H., and Rodriguez, P. (2014). Searching for a unique style in
soccer. \textit{arXiv preprint arXiv:1409.0308}.

\bibitem[Hastie et al. (2009)]{hastie2009elements}
Hastie, T., Tibshirani, R., and Friedman, J. (2009). \textit{The Elements of Statistical
Learning: Data Mining, Inference, and Prediction (Second Edition)}. Springer
Verlag, New York.

\bibitem[Hubert and Arabie (1985)]{hubert1985comparing}
Hubert, L. and Arabie, P. (1985). Comparing partitions. \textit{Journal of classification}, \textbf{2}(1), 193–218.

\bibitem[Hughes and Franks (2005)]{hughes2005analysis}
Hughes, M. and Franks, I. (2005). Analysis of passing sequences, shots and goals in
soccer. \textit{Journal of sports sciences}, \textbf{23}(5), 509–514.


\bibitem[Kaufman and Rousseeuw (1990)]{kaufman1990finding}
Kaufman, L. and Rousseeuw, P. J. (1990). \textit{Finding groups in data: an introduction
to cluster analysis}, volume 344. John Wiley \& Sons.

\bibitem[Koopman and Lit (2015)]{koopman2015dynamic}
Koopman, S. J. and Lit, R. (2015). A dynamic bivariate poisson model for analysing
and forecasting match results in the english premier league. \textit{Journal of the Royal
Statistical Society: Series A (Statistics in Society)}, \textbf{178}(1), 167–186.

\bibitem[Maher (1982)]{maher1982modelling}
Maher, M. J. (1982). Modelling association football scores. \textit{Statistica Neerlandica}, \textbf{36}(3), 109–118.


\bibitem[Molenberghs and Verbeke (2005)]{libropseudoveros}
Molenberghs, G. and Verbeke, G. (2005). Pseudo-likelihood. In Molenberghs, G. and
Verbeke, G., editors, \textit{Models for discrete longitudinal data}, pages 189–202. Springer.

\bibitem[Narizuka et al. (2014)]{narizuka2014statistical}
Narizuka, T., Ken Y., and Yoshihiro Y. (2014). Statistical properties of position-dependent ball-passing networks in football games. \textit{Physica A: Statistical Mechanics and Its Applications} \textbf{412}: 157-168.


\bibitem[Newmark (2010a)]{newman2010networks}
Newman, M. (2010a). Mathematics of networks. In Newman, M., editor, \textit{Networks:
an introduction}, pages 109–167. Oxford university press.

\bibitem[Newmark (2010b)]{newman2010networkscap11}
Newman, M. (2010b). Matrix algorithms and graph partitioning. In Newman, M.,
editor, \textit{Networks: an introduction}, pages 345–394. Oxford university press.





\bibitem[Pe{\~n}a (2014)]{pena2014markovian}
Pe{\~n}a, J. L. (2014). A markovian model for association football possession and its
outcomes. \textit{arXiv preprint arXiv:1403.7993}.

\bibitem[Pe{\~n}a and Navarro (2015)]{pena2015can}
Pe{\~n}a, J. L. and Navarro, R. S. (2015). Who can replace xavi? a passing motif analysis
of football players. \textit{arXiv preprint arXiv:1506.07768}.

\bibitem[Pe{\~n}a and Touchette (2012)]{pena2012network}
Pe{\~n}a, J. L. and Touchette, H. (2012). A network theory analysis of football strategies.
\textit{arXiv preprint arXiv:1206.6904}.

\bibitem[Pina et al. (2017)]{pina2017network}
Pina, T. J., Paulo, A., and Ara{\'u}jo, D. (2017). Network characteristics of successful
performance in association football. a study on the uefa champions league. \textit{Frontiers
in psychology}, \textbf{8}, 1173.



\bibitem[R Core Team (2016)]{Rsoftware}
R Core Team (2016). \textit{R: A Language and Environment for Statistical Computing}.
R Foundation for Statistical Computing, Vienna, Austria. URL \url{https://www.R-project.org/}.



\bibitem[Rand (1971)]{rand1971objective}
Rand, W. M. (1971). Objective criteria for the evaluation of clustering methods.
\textit{Journal of the American Statistical association}, \textbf{66}(336), 846–850.



\bibitem[Santini (2014)]{Santinigoal}
Santini, A. (2014). Quando il calcio dura poco: in serie a non si raggiunge l’ora di gioco, ma non siamo i peggiori. URL http://www.goal.com/it/news/2/serie-a/2014/03/11/4676323/quando-il-calcio-dura-poco-in-serie-a-non-si-raggiunge-lora.

\bibitem[Satterthwaite (1946)]{satterthwaite1946approximate}
Satterthwaite, F. E. (1946). An approximate distribution of estimates of variance
components. \textit{Biometrics bulletin}, \textbf{2}(6), 110–114.

\bibitem[Sokal and Michener (1958)]{sokal1958statistical}
Sokal, R. R. and Michener, C. D. (1958). A statistical method for evaluating system
atic relationship. \textit{University of Kansas science bulletin}, \textbf{28}, 1409–1438.

\bibitem[Varin et al. (2011)]{varin2011overview}
Varin, C., Reid, N., and Firth, D. (2011). An overview of composite likelihood
methods. \textit{Statistica Sinica}, pages 5–42.










%
%
%
%
%
\end{thebibliography}


\end{document}